\documentclass[journal]{IEEEtran}
\usepackage{amsfonts, cite}
\usepackage{amssymb}
\usepackage{color}
\usepackage{subfigure}
\usepackage{scalefnt}
\usepackage{amsmath}
\usepackage{amsfonts}
\usepackage{colordvi}
\usepackage{verbatim}
\usepackage{graphicx}
\usepackage{latexsym}
\usepackage{bm}
\usepackage{cite}
\usepackage{graphicx}
\usepackage{graphics}
\usepackage{times}

\usepackage{algpseudocode}

\newcommand{\Ex}{\mathbb{E}\/}
\newcommand{\Px}{\mathbb{P}\/}
\newcommand{\CN}{\mathcal{CN}}
\newcommand{\LN}{\mathcal{LN}}

\usepackage{paralist}
\defaultleftmargin{1em}{}{}{}

\title{Analyzing Dependent Placements of Small Cells in a Two-Layer Heterogeneous Network with a Rate Coverage Constraint}%

\author{S. Alireza~Banani, Andrew W.~Eckford, Raviraj~S.~Adve%

\thanks{This research was supported by TELUS Canada and the Natural Sciences and Engineering Research Council (NSERC) of Canada.}
\thanks{S. A. Banani and R. S. Adve are with the ~Dept. of Electrical and Computer Engineering, University of Toronto, Toronto, Canada; emails: alireza.banani@utoronto.ca, rsadve@comm.utoronto.ca}%
\thanks{A. W. Eckford is with the Dept. of Electrical Engineering and Computer Science, York University, Toronto, Canada; email: aeckford@yorku.ca}}

\IEEEaftertitletext{\vspace{-1.25\baselineskip}}

\begin{document}
\maketitle \thispagestyle{empty}

\begin{abstract}
We consider the downlink of a two-layer heterogeneous network, comprising macro cells (MCs) and small cells (SCs). The existing literature generally assumes independent placements of the access points (APs) in different layers; in contrast, we analyze a dependent placement where SC APs are placed at locations with poor service from the MC layer. Our goal is to obtain an estimate of the number of SCs required to maintain a target outage rate. Such an analysis is trivial if the MCs are located according to a Poisson point process (PPP), which provides a lower bound on performance. Here, we consider MCs placed on a hexagonal grid, which complements the PPP model by providing an upper bound on performance. We first provide accurate bounds for the average interference within a MC when SCs are not used. Then, by obtaining the outage areas, we estimate the number of SCs required within an MC to overcome outage. If resource allocation amongst SCs is not used, we show that the problem of outage is not solved completely, and the residual outage area depends on whether co-channel or orthogonal SCs are used. Simulations show that a much smaller residual outage area is obtained with orthogonal SCs.


\end{abstract}

\section{Introduction}
\label{sec:introduction}

Heterogeneous networks (HetNets) are being considered as an efficient way to improve overall system rates as well as enhance network coverage~\cite{Ref1,Ref2}. Two-layer HetNets comprise the conventional macrocellular  network, with macro basestations (BSs) forming the first layer, overlaid with diverse sets of small cell (SC) access points (APs) in a higher layer. Crucially, the existing HetNet literature has generally assumed that the SCs are deployed independently of the macrocell (MC) layer~\cite{Ref3}. However, in the cellular scenario, SCs are expected to be installed for range extension and to compensate for holes in MC coverage; the locations of BSs and SC APs cannot, therefore, be treated as independent. The situation considered in this paper is the \emph{joint} placement of MCs and SCs, where both groups of cells are under the control of a single service provider.

\subsection{Related Work}

There is much existing work on the structure of HetNets. A widely-used analytical model is the homogeneous Poisson point process (PPP), in which the locations of nodes (BSs or SC APs) in each layer \emph{independently} follow a PPP of varying densities; a superposition of independent layers models the entire network~\cite{andrews10,Ref8,Ref9,Ref10,Ref11_new,Ref12_new,Ref13_new}. The independence of node placement amongst layers provides analytical tractability and, coupled with tools from stochastic geometry, leads to useful signal-to-interference-plus noise ratio (SINR) and/or rate expressions. Techniques from stochastic geometry~\cite{Ref4,Ref5,Ref6,Ref19} and large deviation theory~\cite{Ref7} have also been used to analyze the resulting irregularities in cell deployments and interference in the network.

In considering the (non-random) placement of SCs, some recent work has considered optimization; however, this research often focuses on indoor deployment, perhaps because such deployments can be carefully controlled. On the subject of joint macro-femto deployment scenarios, the work in~\cite{calin10} presents an overview of joint deployments, where simulations are used to quantify the benefits of offloading traffic from the MC. These benefits are a key component of any business case related to the introduction of femtocells~\cite{lin11}.

Since independent location models do not incorporate dependencies between layers, they do not accurately capture the objective of SC deployments in enhancing the coverage or rate of the network. Only recently has a location-dependent model based on a PPP model for BS locations been proposed for heterogeneous networks~\cite{Ref11}. The proposed model comprises multiple tiers including a homogenous PPP in the first tier and a non-homogenous PPP restricted to the edges of Voronoi cells of tier 1, on the second tier (since, when ignoring shadowing, these edges and vertices are the locations with poorest coverage by the MC BSs). However, while the proposed model provides trade-offs between accuracy and tractability, no analysis is given and the results are based on simulations. We also note that the problem of dependent placement of layers in a HetNet is different from the problems of user association and dependent connectivity~\cite{Ref14_new} or priority scheduling and handover~\cite{Ref15_new,Ref16_new} between different layers of a network.

\subsection{Our Contributions}

As is clear from our literature review, there has been little work on dependent placement of APs, and even less in terms of analysis. In this regard, while based on some simplifying assumptions and approximations (confirmed via simulations), our analysis provides both insights and design tools for network planning. We consider the downlink of a two-layer heterogeneous network where we place SCs at locations with poor rate coverage from the macro BSs. Our motivation for using a SC layer is to enhance coverage probability in the network, not to offload traffic off the BSs. We estimate the number of SCs required to achieve a chosen coverage metric; crucially, and unlike most previous works, we include shadowing within our channel model. We consider frequency-reuse 1 amongst BSs, i.e., all BSs share all frequencies and intercell interference from surrounding BSs must be accounted for.

We begin with analyzing the coverage holes in the macrocell layer. To do so, there are two popular models for the BS locations: the PPP model, which has been shown to provide a lower bound on system performance (e.g., coverage probability), and the hexagonal lattice model, which provides an upper bound~\cite{guo2013spatial,andrews2011tractable}. If the locations of the BSs followed a PPP (and we ignored shadowing), the required analysis would be trivial - outage area of a single-layer PPP is well known~\cite{Ref9}. Simply dividing this outage area with the expected coverage area of a single SC provides the required estimate. In our work, we consider the BSs to be located on a hexagonal lattice; our results are new in providing some analytic tractability for hexagonal lattice networks.

Our first major contribution is to obtain the average area of a MC in outage with respect to a chosen rate threshold. Outage is due to path loss, shadowing and inter-cell interference. We do this by developing simple, tight, lower and upper bounds on the average total interference in a single-layer hexagonal reuse-1 network. The bounds are obtained from curve-fitting and are presented in the form of third order polynomial functions of the normalized distance from the centre of the MC (the normalization is to the radius of the MC). The proposed bounds are used to obtain analytical expressions for various network parameters, such as SINR and/or rate outage probability (ROP), as functions of normalized distance to the centre of the MC. Assuming isolated SCs (SCs that do not receive interference from other SCs or the MC) are placed in outage areas, we then provide analytical expressions for the average number of SCs required to completely cover the ares in outage.

Our second major contribution is to show, through simulations, the placement of SCs within a chosen macrocell for given realizations of shadowing; service providers may use our method in a network design tool. We also show that without resource allocation amongst SCs, the problem of outage is not solved completely, since the interference from nearby SCs now play a role and can result in a residual outage area. The residual outage area is evaluated via simulations. We consider the cases of orthogonal and co-channel frequency allocation across the MC and SC layers. We use simulations to show that, despite incorporating small cells, using a non-orthogonal allocation does not, in fact, reduce the total outage area.

Finally, while all the analysis in this paper is for frequency reuse-1 in the macrocell layer, we compare the cases of reuse-1 and reuse-7 via simulations. Using reuse-7 represents a trade-off between reduced interference and reduced available frequency resources. For a given constraint on required rate coverage, we show that a reuse-7 network suffers from a larger residual outage area at moderate to high values of MC radius.

\subsection{Organization and Notation}

The rest of the paper is organized as follows: Section~\ref{sec:Downlink System Model} describes the downlink system model while the bounds on the average total interference within a MC are derived in Section~\ref{sec:bounds}. These bounds are needed to formulate the SIR and the ROP within a MC which, in Section~\ref{sec:NoOfSCs}, are used to obtain the average required number of SCs. Supporting simulation results, illustrating the efficacy of the analysis and important aspects of the problem at hand, are presented in Section~\ref{sec:Simulations}. Finally, Section~\ref{sec:conc} wraps up the paper with a  review and some conclusions.

The notation used is conventional: matrices are represented using bold upper case and vectors using bold lower case letters; $(\cdot)^{H}$ , and $(\cdot)^{T}$ denote the conjugate transpose, and transpose, respectively. $a \sim \mathcal{N}(\mu,\sigma^2)$ or $a \sim \CN(\mu,\sigma^2)$ denote, respectively, a real or complex Gaussian random variable with mean $\mu$ and variance $\sigma^2$ while $X\sim\LN(\mu_{x},\sigma_x)$ represents a log-normal random variable where $\ln(X) \sim \mathcal{N}(\mu_{x},\sigma_x^2)$. $Q(x)$ represents the standard \emph{Q}-function, the area under the tail of a standard Gaussian distribution. $\Px\{\cdot\}$ denotes the probability of an event and $\Ex\{\cdot\}$ denotes expectation. Finally, Table~\ref{tab:symbols} lists the important symbols used throughout the paper.

\begin{table*}[t]
\caption{List of Important Symbols}
 \label{tab:symbols}
 \centering
 \begin{tabular}{|c|c||c|c|}
   \hline
   Symbol & Description & Symbol & Description\\
 \hline \hline
 $W$ & total available bandwidth in the network & $A_\textmd{MC}$ & the area of a single macrocell \\ \hline
 $W_{0}$ & bandwidth allocated to a user & $r_\textmd{MC}$ & circumscribed radius of a hexagonal MC \\ \hline
 $M$ & number of frequency slots & $A_\textmd{SC}$ & the coverage area of a single SC \\ \hline
 $h_{k}$ & channel power between the $k$-th BS and a user & $r_\textmd{SC}$ & circumscribed radius of a SC  \\ \hline
 $\tilde{g}_{k}$ & Rayleigh fading from $k$-th BS to the user & $r_{k}$ & distance of user to the $k$-th BS \\ \hline
 $L_{k}$ & $\LN$ random variable representing the shadowing & $\sigma_M^2$ & MC transmit power \\ \hline
 $\sigma_{L}$ & shadowing standard deviation & $\sigma^{2}_{n}$ & power of thermal noise \\ \hline
 $\textmd{PL}(\,\cdot \,)$ & pathloss (in dB) & $a_i$ & $i$-th coefficient in the polynomial curve fitting \\ \hline
 $\alpha$ & path loss exponent  & $P$ & the order used in polynomial curve fitting \\ \hline
 $r_{\textmd{ref}}$ & close-in reference path loss distance & $N$ & the average number of required SCs \\ \hline
 $I(\, \cdot \,)$ & total interference power from surrounding BSs & $\texttt{A}_{1}$ & \begin{tabular}{@{}c@{}} area of a hexagonal region within a MC where users are \\ associated with the BS located at the center of the cell \end{tabular} \\ \hline
 $C$ & achievable spectral efficiency (b/s/Hz) & $\texttt{A}_{2}$ & \begin{tabular}{@{}c@{}} area of a rectangular region between two MCs that users \\ may receive service from either of the two nearby BSs \end{tabular} \\ \hline
 $C_0$ & required spectral efficiency (b/s/Hz) & $\texttt{A}_{3}$ & \begin{tabular}{@{}c@{}} area of a triangular region near the corner of the MC where \\ users may communicate with any of the three nearby BSs \end{tabular} \\ \hline
 $R$ & per-user data rate (b/s) & $g$ & parameter relating the three regions $\texttt{A}_{1}$, $\texttt{A}_{2}$ and $\texttt{A}_{3}$ \\ \hline
 $R_0$ & target data rate (b/s) & $\gamma_{g}$ & a parameter relating $g$ to the inscribed circle of the MC \\ \hline
 $\Gamma$ & SNR gap to capacity & $\delta^{(i)}$ & the fractional outage area within region $\texttt{A}_{i}$ \\ \hline
 \end{tabular}
 \end{table*}

\section{Downlink System Model}
\label{sec:Downlink System Model}

The system model at hand comprises MC BSs, SC APs and users distributed in the network. We begin by listing the assumptions we  make about these components of the network.
\begin{itemize}

\item Users:
\begin{itemize}
	\item Users are uniformly distributed within the network.
	\item We assume a total available bandwidth of $W~\textmd{Hz}$ equally divided in $M$ frequency slots only one of which is assigned to a user, i.e., each active user is allocated the same bandwidth of $W_{0}=W/M~\textmd{Hz}$.
	 \item If SCs are not present, each user attempts to connect to the BS with the strongest downlink signal (averaged over the small-scale fading).
\end{itemize}

\item MCs and BSs:
\begin{itemize}
\item The MCs form a reuse-1 interference-limited network and we ignore thermal noise~\cite{Ref13}.
 \item We assume the network is fully loaded, i.e., all $M$ frequency slots have been allocated.
 \item The BSs in the network are assumed to be placed on a hexagonal lattice where each hexagon has a circumscribed radius of $r_\textmd{MC}$.
 \item All BSs transmit at a power level of $\sigma_M^2$.
\end{itemize}

\item Interaction between MCs and SCs:
\begin{itemize}
 \item While each MC covers an area of $A_\textmd{MC}$, each SC is assumed to provide service in an area of $A_\textmd{SC}$. As a crucial simplification of the geometry involved, we assume that the service area of a SC is also a hexagon - this allows us to place SCs within a MC in a hex-tile format. (Figure~\ref{new_Fig_8} illustrates this hex-tile arrangement).
 \item We assume a piecewise constant and uncorrelated shadowing model. This is akin to the popular block fading model for small-scale fading - the shadowing is assumed constant over the coverage area of a single SC, i.e., shadowing is considered to be constant within an area of $A_{\textmd{SC}}$ and the shadowing is independent from one area of size $A_{\textmd{SC}}$ to another.
\end{itemize}

\end{itemize}

\subsection{Macrocell Layer}

\begin{figure}[t]
\begin{center}
\includegraphics[width=0.3\textwidth]{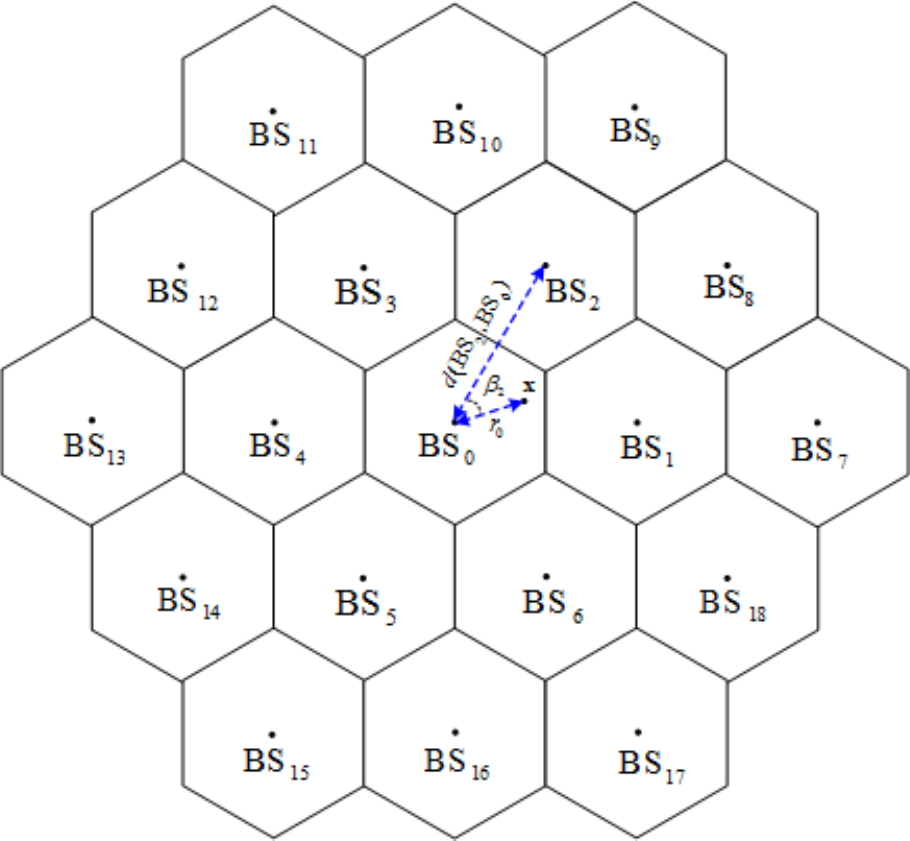}
\caption{Schematic of a MC in a hexagonal network with two tiers of surrounding interferers in a resue-1 network}
\label{Fig_1}
\end{center}
\end{figure}

Consistent with our assumptions, consider a large, resue-1, HetNet with BSs (first layer) located at the centers of the hexagons as shown in Fig.~\ref{Fig_1}. For such a symmetric model, with MCs covering the entire plane, it is adequate to concentrate on a single cell. The $M$ active users in the cell are each allocated a fixed bandwidth of $W_{0}=W/M~\textmd{Hz}$. Communications to these users is limited by the interference from the surrounding macros-BSs. We assume that users connect to the BS with the strongest downlink signal. Our channel model accounts for path loss, shadowing and small-scale fading. Therefore, due to different realizations of shadowing, users at the edges of a cell may connect to a BS that is not physically the closest.

Let $h_{k}, k=0,1,\cdots,$ denote the channel powers between the surrounding BSs and a user at location $\textbf{x}$, within the MC under consideration. We denote as $h_0$ the channel power from the BS at the center of the cell under consideration while $h_k, k = 1,\dots$ represent the channel powers from the surrounding BSs to a user. Let $r_{k};k=0,1,\cdots,$ represent the corresponding distances from the BSs to the user; here $r_0$ represents the distance of user to the BS in the cell under consideration, and $r_k, k = 1, \dots$ represent the distances to the surrounding BSs. The channel powers $h_{k}$ can be modelled as
\begin{equation} \label{eq_1}
 h_{k}(r_{k})=|\bar{g}_{k}|^2 10^{-(PL(r_{k})+L_{k})/10}; \hspace*{0.2in} k=0,1,\cdots,
\end{equation}
where $\tilde{g}_{k} \sim \CN(0,1)$ represents the normalized complex channel gain, reflecting small-scale Rayleigh fading from the $k$-th BS to the user, which is independent from $\bar{g}_{k'}; k' \neq k$, and where $L_{k}$ is a random variable representing the shadowing; $L_{k}\sim \mathcal{N}(0,\sigma_{L})$ is modeled as log-normal fading with standard deviation $\sigma_{L}$ expressed in dB (the value of $\sigma_{L}$ depends on the environment~\cite{Ref12}) and $\textmd{PL}(r_{k})$ represents the pathloss (in dB) over the distance $r_{k}$ expressed as
\begin{equation} \label{eq_2}
\textmd{PL}(r_{k})=10\alpha \log(\max\{r_{\textmd{ref}},r_{k}\}).
\end{equation}
Here, $\alpha >2$ is the path loss exponent (PLE) and $r_{\textmd{ref}}$ is the close-in reference path loss distance which is determined from measurements close to the transmitter.

The SINR received at a location $\textbf{x}$, from the BS in the cell under consideration, can be expressed as
\begin{equation} \label{eq_3}
\begin{split}
\mathtt{SINR}_{0} & =\frac{\sigma^{2}_{M}h_{0}(r_{0})}
{\sigma^{2}_{n}+I(\textbf{x})}=\frac{\sigma^{2}_{M}h_{0}(r_{0})}
{\sigma^{2}_{n}+\sum_{k=1}^{\infty}
\sigma^{2}_{M}h_{k}(r_{k})} \\
& \simeq \frac{\sigma^{2}_{M}h_{0}(r_{0})}
{\sum_{k=1}^{\infty}
\sigma^{2}_{M}h_{k}(r_{k})} ,
\end{split}
\end{equation}
where $\sigma^{2}_{M}$ and $\sigma^{2}_{n}$ represent the power of the transmitted signals and thermal noise, respectively, and $I(\textbf{x})$ represents the total interference power from surrounding BSs. The thermal noise is ignored in the final expression. The subscript in $\mathtt{SINR}_{0}$ indicates that this is the SINR that a user at position $\textbf{x}$ could achieve from the BS at the centre of the MC under consideration, which is not necessarily the strongest received signal from nearby BSs. (In Section \ref{sec:NoOfSCs}, we exploit the hexagonal geometry of the network to determine which BS provides the strongest signal.)

The user connects to the BS providing the largest SINR (averaged over the small-scale fading). With a bandwidth of $W_{0}=W/M$ available to each user, the per-user data rate (in b/s) of $R = W_{0}\log_{2}(1+\mathtt{SINR}_{\max}/\Gamma)$ is achievable at each point in the MC; where $\Gamma$ is the SNR gap to capacity~\cite{Ref_e2}, and $\mathtt{SINR}_{\max}$ denotes the maximum of SINR from nearby BSs. The rate outage probability (ROP) at each point of a MC is defined as the probability that a particular user can not achieve a target rate $R_0$:
\begin{equation} \label{eq_7}
\begin{split}
\Px\{R<R_{0}\}= & \Px\left\{W_0 \log_{2}\left(1+\frac{\mathtt{SINR}_{\max}}{\Gamma}\right) <
R_{0}\right\}, \\
= & \Px\left\{\mathtt{SINR}_{\max}<\Gamma(2^{C_{0}}-1)\right\}= \Px\{C<C_{0}\},
\end{split}
\end{equation}
where we define $C = R/W_0$ and $C_0 = R_0/W_0$, as the achievable and required spectral efficiencies (in b/s/Hz). A point in a cell is said to be in outage if the ROP at that point is greater than a threshold $\eta$, i.e., $\Px\{C<C_{0}\} \geq \eta$.

\subsection{Small Cell Layer}
This section explains the possible configurations of the SC layer; specifically the possible frequency allocation schemes used in the SC layer and the associated interference patterns. Since users are assumed to be uniformly distributed, when required, the frequency allocation is largely in proportion to service area.
\begin{itemize}
 \item \emph{Co-channel SCs} share all $M$ frequencies with the serving MC; as a consequence co-channel SCs experience interference both from other SCs as well as the BSs. The BS can serve $M$ users even with SCs operating.
 \item \emph{Orthogonal SCs} use a different frequency band from the one used by the BS, but all SCs share the same frequencies. Specifically, since each SC serves an area of $A_\textmd{SC}$ and a MC an area of $A_\textmd{MC}$, a fraction $A_\textmd{SC}/A_\textmd{MC}$ of the $M$ frequencies are allocated to the small cell layer with the remaining allocated to the MC.

     Since an orthogonal allocation is used, the BSs and SCs do not interact; users in the SCs experience interference from nearby SCs only. The penalty is that the BS layer is left with a total available bandwidth of $W(1-A_\textmd{SC}/A_\textmd{MC})$ serving $M(1-A_\textmd{SC}/A_\textmd{MC})$ users.

 \item \emph{Isolated SCs} use an orthogonal resource allocation (e.g., different frequency bands) among SCs and the BS. Therefore, users do not experience any interference from other SCs or the BSs. Both the BS and SCs have fewer frequency slots available. With $N$ SCs, a fraction of $N A_{\textmd{SC}}/A_{\textmd{MC}}$ of the total available bandwidth to a MC is dedicated to SCs. Again, the penalty for the reduced interference is that the BS can serve only $M(1-NA_\textmd{SC}/A_\textmd{MC})$ users.
\end{itemize}
It is worth noting that in all cases a fixed bandwidth of $W_{0}=W/M$ is allocated to each user, with or without the use of SCs. The difference is in the frequency slots available, and hence, the number of users that can be simultaneously served by the BS.

Since users connected to isolated SCs do not suffer interference, given enough frequency resources, outage areas within a MC (in outage when using BSs only) can be covered completely by SCs, leaving no residual outages in the network. In Section~\ref{sec:NoOfSCs}, we first provide an analytical expression for the average required number of such isolated SCs within a MC. However, if using orthogonal or co-channel SCs, interference implies that the problem of outage is not solved completely, and the network experiences a residual outage area. The residual outage area depends heavily on whether co-channel or orthogonal SCs are used. In Section~\ref{sec:Simulations}, we evaluate and compare the average residual outage area for these cases.

\section{Bounds on Average Total Interference in a Reuse-1 Network}
\label{sec:bounds}

As indicated earlier, for our purposes, it is sufficient to consider a single MC within the network; the BS at the centre of this MC is denoted $\textmd{BS}_0$. Let $r$ denote the distance of a user at location $\textbf{x}$, within the MC, to the base station providing the greatest SINR. For now, we will assume that the user connects to the $\textmd{BS}_0$, i.e., $r=r_0$. Since the locations of the interfering BSs are deterministic and known, one may try to express $I(\textbf{x})$ in~\eqref{eq_3} as a function of $r$ by relating the distances $r_{k}$ to $r$  via the triangle cosine rule as
\begin{equation} \label{eq_9}
\begin{split}
r_{k}^2 = r^2 + d(\textmd{BS}_{k},\textmd{BS}_{0})^2 & -2r_{k}d(\textmd{BS}_{k},\textmd{BS}_{0})\cos \beta_{k}, \\
& ~~~~~~~~~~~~~~~~~ k = 1, \dots,
\end{split}
\end{equation}
where $d(\textmd{BS}_{k},\textmd{BS}_{0})$ denotes the distance between $\textmd{BS}_{k}$ and $\textmd{BS}_{0}$, and $\beta_{k}$ is the angle between the lines connecting $\textmd{BS}_{k}$ and $\textmd{BS}_{0}$, and the line connecting $\textbf{x}$ and $\textmd{BS}_{0}$ (for example, $d(\textmd{BS}_{2},\textmd{BS}_{0})$ and $\beta_{2}$ are illustrated in Fig.~\ref{Fig_1}). However, using~\eqref{eq_9} leads to a very complicated expression for $I(\textbf{x})$ and the resulting analysis is intractable.

As an alternative, we note that the interference $I(\textbf{x}) = \sum_{k=1}^{\infty} \sigma^{2}_{M} |\bar{g}_{k}|^2 r_{k}^{-\alpha} z_{k}$ is a linear combination of independent log-normal random variables. Here $z_{k} =10^{-L_{k} /10}, k=1,\cdots$ are independent lognormal RVs as $z_{k} \sim \LN(\mu_{z}=0, \sigma _{z} =(0.1\ln 10) \sigma _{L})$. One may use the moment matching approximation (MMA) approach~\cite{pratesi2006generalized} to approximate the interference by a single lognormal random variable. However, it has been shown that the accuracy of the MMA is questionable if the terms in the sum have different means and variances~\cite{mehta2007approximating,di2009further}. In our network model, since the distances $r_{k},k=1,\cdots$ are different; the terms therefore have different means. Consequently, using a log-normal approximation for the interference based on the MMA is inappropriate.

To move forward, we obtain accurate lower and upper bounds on the average total interference. Importantly, the bounds are very simple in the sense that they are approximated in the form of polynomials in the normalized distance $r/r_\textmd{MC}$. Within a hexagonal MC, the worst signal received from the BS at the centre of the cell, averaged over small scale and log-normal fading, is at one of the six corners; this worst case average signal power is $\sigma^{2}_{M}r^{-\alpha}_{\textmd{MC}}\exp(\sigma^{2}_{z}/2)$ with $\sigma_{z}=(0.1\ln10)\sigma_{L}$. Let $\bar{I}^{\,(avg)}(\textbf{x})$ denote the total interference power, averaged over shadowing, normalized to this term, i.e.,
\begin{equation} \label{eq_9_new}
\begin{split}
\bar{I}^{\,(avg)}(\textbf{x})=&\frac{1}{\sigma^{2}_{M}r^{-\alpha}_{\textmd{MC}}
\exp(\sigma^{2}_{z}/2)} \Ex\{I(\textbf{x})\} \\
=& \frac{1}{\sigma^{2}_{M}r^{-\alpha}_{\textmd{MC}}\exp(\sigma^{2}_{z}/2)} \sum_{k=1}^{\infty} \sigma^{2}_{M} \Ex\{h_{k}(r_{k})\} \\
= & \sum_{k=1}^{\infty} \left( \frac{r_{k}}{r_{\textmd{MC}}} \right) ^{-\alpha},
\end{split}
\end{equation}
where the expectation is over the shadowing. The normalization makes the average interference independent of the BSs transmit power ($\sigma^{2}_{M}$), MC radius ($r_{\textmd{MC}}$), and the variance of log-normal shadowing ($\sigma^{2}_{L}$). As a result, the average normalized interference $\bar{I}^{\,(avg)}(\textbf{x})$ is only a function of the PLE, $\alpha$ (and, clearly, the location \textbf{x}). Given the geometry involved (see Fig.~\ref{Fig_1}), we can state a few properties of $\bar{I}^{\,(avg)}(\textbf{x})$, which we emphasize is normalized over the small-scale fading and shadowing:
\begin{itemize}
 \item For a fixed distance, $r$, from the central BS, $\textrm{BS}_0$, the average interference is periodic in angle with period of $\Delta\theta_0 = \pi/6$ (see Fig.~\ref{Fig_1}). It is therefore sufficient to consider the wedge $0 \leq \theta_0 \leq \pi/6$ in our analysis (such as the wedge shown in Fig.~\ref{Fig_2_new}(b)).
 \item Within the wedge, the \emph{maximum} average interference power is along the horizontal ($\theta_0 = 0$) segment connecting $\textmd{BS}_0$ and $\textmd{BS}_1$ (see Fig.~\ref{Fig_2_new}(a)) - along this segment a user would be closest to the dominant interference source, $\textmd{BS}_1$.
 \item Within the wedge, the \emph{minimum} average interference power is along the line defined by $\theta_0 = \pi/6$. Along this segment a user would be furthest from the dominant interference sources.
\end{itemize}

\begin{figure}[t]
\begin{center}
\includegraphics[width=0.5\textwidth]{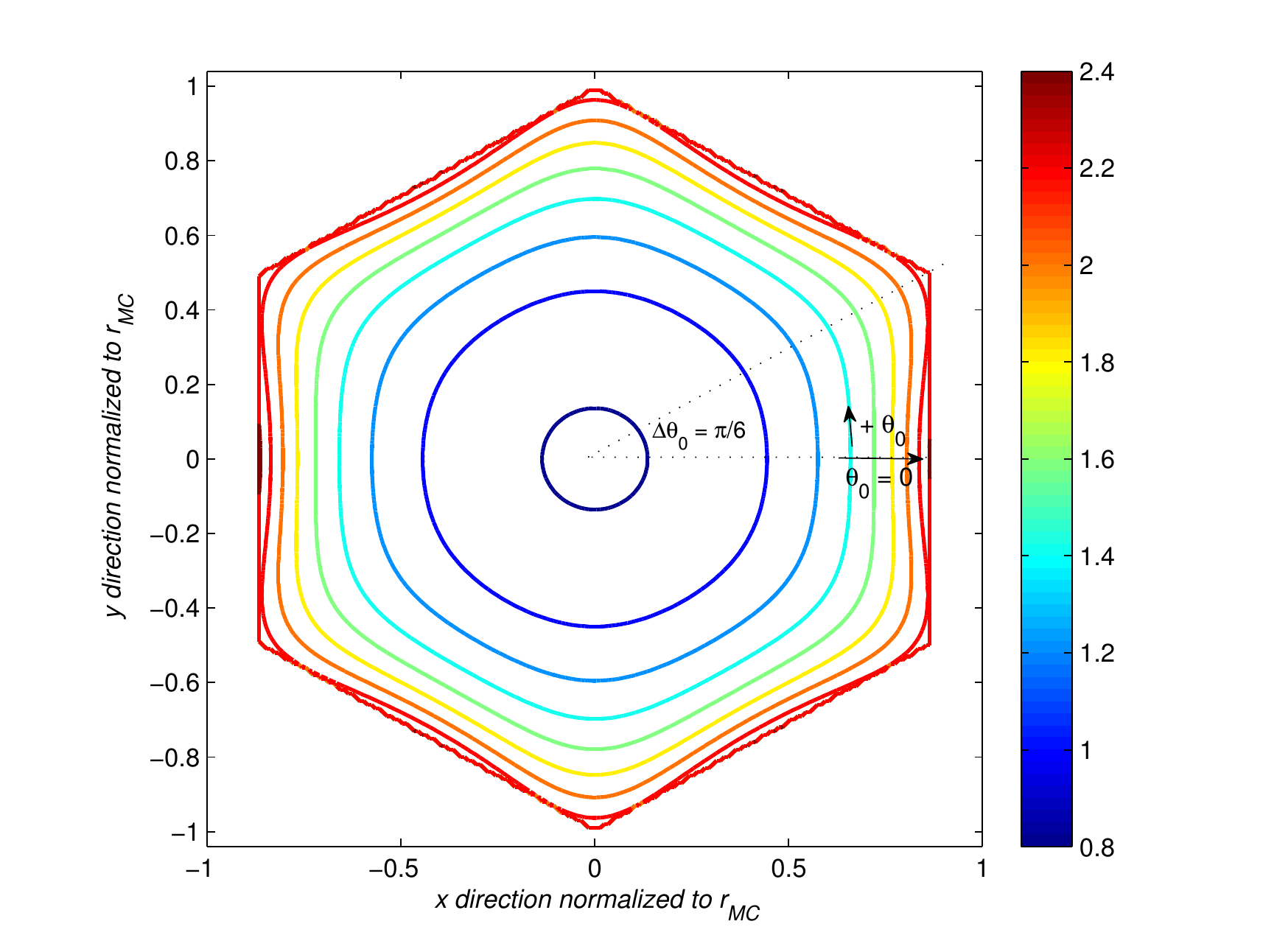}
\caption{Contour plot of the normalized average total interference within a MC of a reuse-1 network with the consideration of two layers of interference from surrounding BSs ($\alpha = 4$).}
\label{Fig_2}
\end{center}
\end{figure}

Figure~\ref{Fig_2} presents contour plots of $\bar{I}^{\,(avg)}(\textbf{x})$ obtained via simulations within a MC for the case of $\alpha=4$. Since the first two tiers of interference make the only significant contributions to the overall received interference, the simulation includes the 18 interfering MCs in the first two tiers: 6 in the first tier and 12 in the second tier\footnote{In our analysis, the interference from BSs further away proved to be negligible. Including more interfering tiers would not change the approach developed here.}. All the distances are normalized to the cell radius $r_{\textmd{MC}}$. Figure~\ref{Fig_2} verifies the dependency of $\bar{I}^{\,(avg)}(\textbf{x})$ on the distance and the angle of the point of interest within a cell; it also confirms that, as expected from Fig.~\ref{Fig_1}, $\bar{I}^{\,(avg)}(\textbf{x})$ is periodic in the angle $\theta_{0}$ with the period $\Delta\theta_{0}=\pi/6$.

\begin{figure}[t]
\begin{center}
\includegraphics[width=0.5\textwidth]{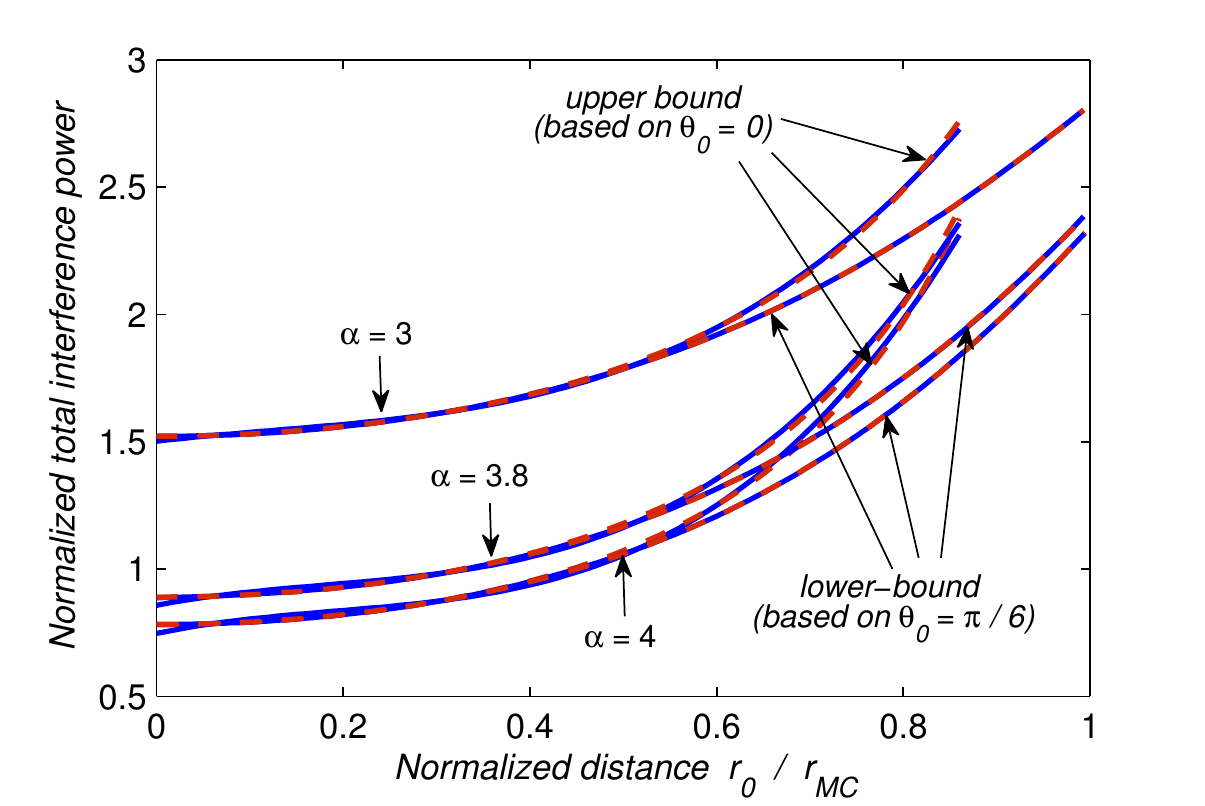}
\caption{Fitting the lower and upper bounds on $\bar{I}^{avg}$ using $3^{\textrm{rd}}$ order polynomial curve fitting, for $\alpha=3$, $\alpha=3.8$ and $\alpha=4$}
\label{Fig_3}
\end{center}
\end{figure}

Consider the interval $0\leq \theta_{0} \leq \pi/6$ and let $\bar{r}=r/r_{\textmd{MC}}$ denote the normalized distance. We can obtain a lower bound and an upper bound on the average total interference power by providing an analytical formulation for $\bar{I}^{\,(avg)}$ at $\theta_{0}=\pi/6$ and $\theta_{0}=0$, respectively. While a closed form expression is intractable, we use an alternative approach: for a given choice of PLE $\alpha$, we evaluate the interference numerically and then use a polynomial fit, as a function of the normalized distance $\bar{r}$. Denoting as $\bar{I}_l^{(avg)}(\bar{r})$ and $\bar{I}_u^{(avg)}(\bar{r})$ the lower and upper bounds (the interference evaluated along the $\theta_0 =\pi/6$ and $\theta_0 = 0$ lines), we express these functions as:
\begin{equation} \label{eq_10}
\hat{\bar{I}}_{l}^{(avg)}(\bar{r})=\sum_{i=0}^{P}a^{\,l}_i\,\bar{r}^{P-i},
\end{equation}
\begin{equation} \label{eq_11}
\hat{\bar{I}}_{u}^{(avg)}(\bar{r})=\sum_{i=0}^{P}a^u_i\,\bar{r}^{P-i},
\end{equation}
where $a^u_i$ or $a^l_i$, $i=1,\cdots,P$ are, respectively, the coefficients found by curve fitting the numerical values obtained from simulations (the subscripts/superscripts '$l$' and '$u$' in~\eqref{eq_10}-\eqref{eq_11} refer to the lower and upper bound, respectively). Figure~\ref{Fig_3} illustrates the resultant approximations of the lower and upper bounds on $\bar{I}^{\,(avg)}$ using a $3^{\textrm{rd}}$ order polynomial, for the three PLEs of $\alpha=3$, $\alpha=3.8$ and $\alpha=4$. The associated coefficients are given in Table~\ref{Tab1}. The dotted lines in the figure are the exact results from simulations. As is clear, a $3^{\textrm{rd}}$ order polynomial is adequate to describe the bounds of interest, though, if required, an even more accurate fit is possible with a higher polynomial degree.
\begin{table}
\caption{The coefficients, $a^{l/u}_i; i=0,\cdots,P$, of the polynomials for the lower and upper bounds on $\bar{I}^{avg}$ using 3-rd order polynomial curve fitting for PLEs $\alpha=3$, $\alpha=3.8$ and $\alpha=4$}
 \label{Tab1}
 \begin{center}
 \begin{tabular}{|c||c|c|c|c|}
   \hline
   Coefficients & $a_0$ & $a_1$ & $a_2$ & $a_3$\\
 \hline \hline
   Upper bound ; $\alpha=4$   & $4.2482$ &        $-2.4301$ &        $0.7687$ &       $0.7469$ \\  \hline
   Lower bound ; $\alpha=4$   & $1.1021$ &        $ 0.3650$ &          $0.1019$ &      $0.7784$ \\  \hline

   Upper bound ; $\alpha=3.8$  & $3.8440$ &    $ -2.0756$ &        $ 0.6881$ &       $0.8581$ \\  \hline
   Lower bound ; $\alpha=3.8$  & $0.5138$ &  $ 0.7843$ &      $0.0109$ &  $1.5217$ \\  \hline

   Upper bound ; $\alpha=3$  & $2.4110$ &    $-0.8962$ &        $0.4137$ &       $1.5024$ \\  \hline
   Lower bound ; $\alpha=3$  & $0.5138$ &  $ 0.7843$ &      $0.0109$ &  $1.5217$ \\  \hline
 \end{tabular}
 \end{center}
 \end{table}

Finally, we can formulate a lower and an upper bound on the average total interference power as a function of distance $r$ as
\begin{equation} \label{eq_12}
\hat{I}_{\,l/u}^{\,(avg)}(r)\simeq \sigma^{2}_{M}\,r^{-\alpha}_{\textmd{MC}}\exp(\sigma^{2}_{z}/2)\,
\hat{\bar{I}}^{\,(avg)}_{\,l/u}(r/r_{\textmd{MC}}).
\end{equation}

Note that since $\bar{I}^{\,(avg)}$ is periodic in the angle $\theta_{0}$, the result in~\eqref{eq_12}, in fact, provides lower and upper bounds on the average interference in the entire macrocell. It is worth emphasizing that, for a given choice of PLE, $\alpha$, the polynomial fit for the lower and upper bounds needs to be obtained only once.

\section{Average Number of Required SCs}
\label{sec:NoOfSCs}
In the previous section we developed expressions for bounds on the average total interference (averaged over shadowing and small-scale fading) at any point in the region of interest. We are now able to develop our result for the average number of SCs required to provide adequate service in the coverage holes.

\subsection{Network geometry}

\begin{figure}[t]
 \begin{center}
 \subfigure[]{\includegraphics[width = .35\textwidth]{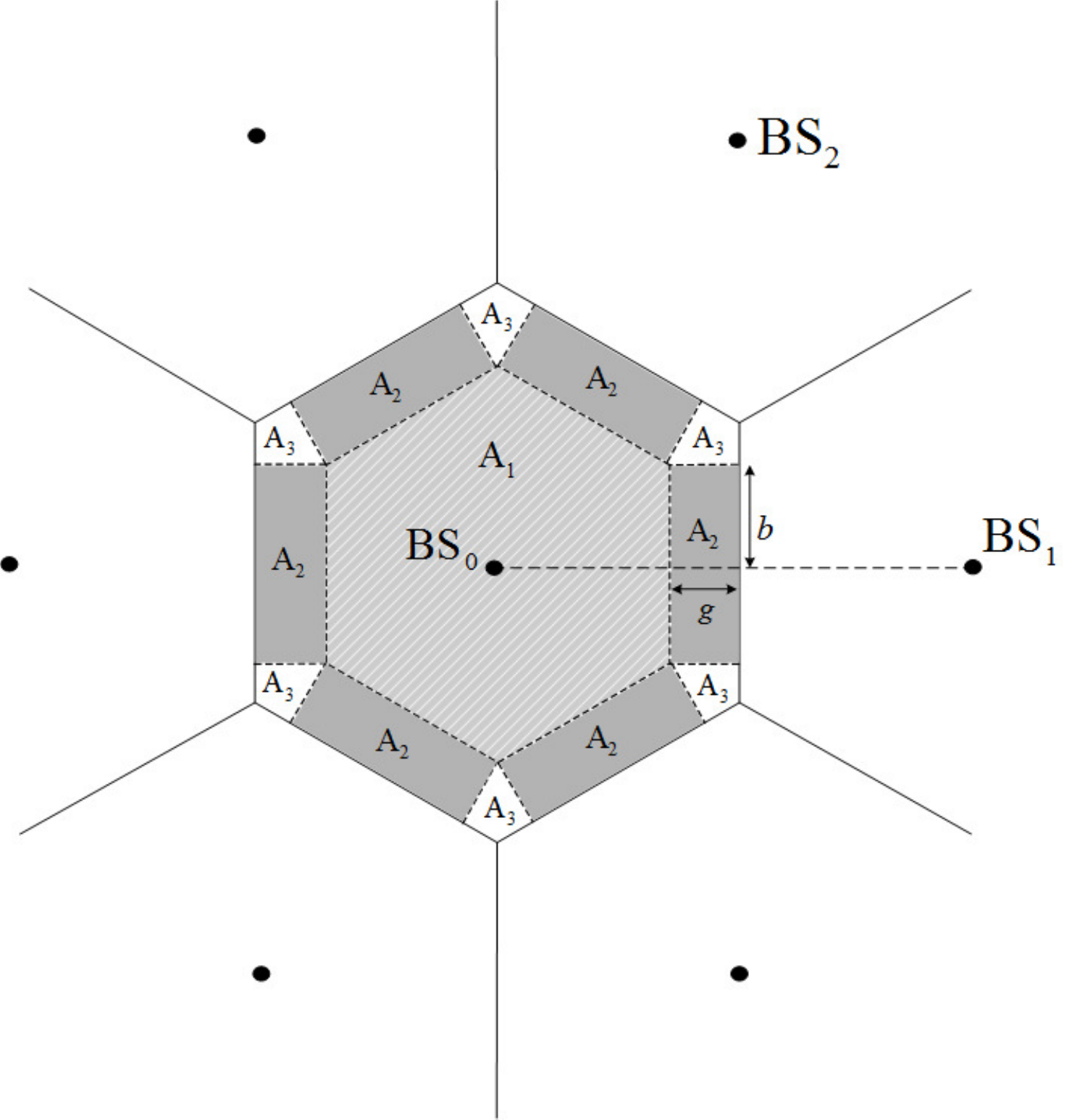}}
 \subfigure[]{\includegraphics[width = .25\textwidth]{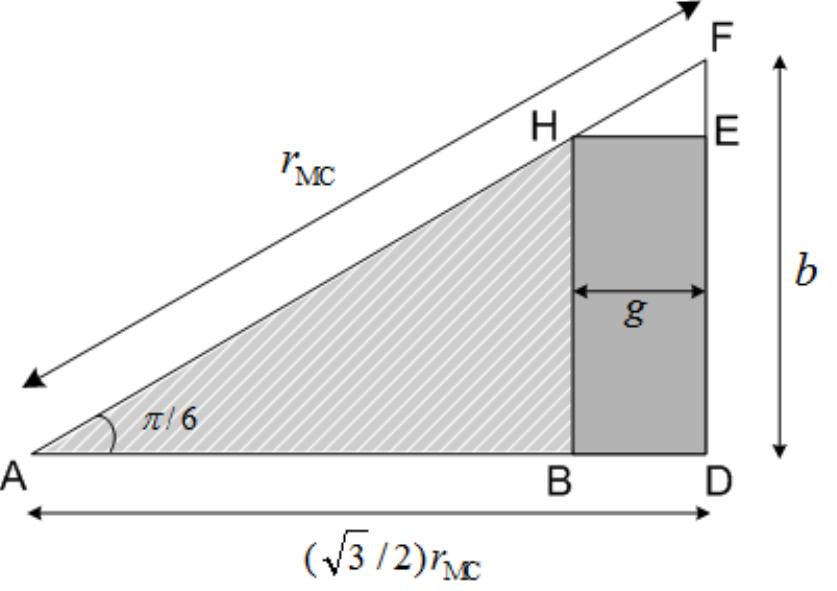}}
 \caption{(a) Separation of MC into regions $A_{1}$, $A_{2}$, and $A_{3}$; (b) Schematic of one $\pi/6$ sector of the MC area}
 \label{Fig_2_new}
 \end{center}
 \end{figure}

In our system model, users are associated with the strongest BS, not necessarily the closest one. As depicted in Fig.~\ref{Fig_2_new}, we develop a geometric approximation for this condition by dividing the area of each MC into three regions:
\begin{itemize}
\item Users within region $\texttt{A}_{1}$, a hexagonal region with circumscribed radius $r_{A_{1}}$, are always associated with the BS located at the center of the cell;
\item Users in region $\texttt{A}_{2}$, at the boundary between two MCs, may receive service from either of the two nearby BSs; and, finally,
\item Users in $\texttt{A}_{3}$, near the corner of the hexagonal tessellation, may communicate with any of the three nearby BSs.
\end{itemize}
The areas $\texttt{A}_{2}$ and $\texttt{A}_{3}$, can be interpreted as guard regions for $\texttt{A}_{1}$; as shown in Fig.~\ref{Fig_2_new}(b), all three regions are defined by a single parameter, $g$. This parameter is picked to ensure that, with probability close to 1, users in region $\texttt{A}_1$ would have picked the BS at the center of the hexagon as having the strongest signal. For simplicity, we relate $g$ to the radius of the inscribed circle of the MC as $g=\gamma_{g}(\sqrt{3}/2)r_{\textmd{MC}}$ ~where $0\leq \gamma_{g} \leq 1$ is now a flexible parameter chosen by the designer. Now, with $r_{\texttt{A}_{1}}=(1-\gamma_{g})r_{\textmd{MC}}$ it follows that
\begin{equation} \label{eq_4}
\begin{split}
A_{1}= & (3\sqrt{3}/2)(1-\gamma_{g})^{2} r^{2}_{\textmd{MC}}, \\
A_{2}= & 3\sqrt{3}\gamma_{g}(1-\gamma_{g})r^{2}_{\textmd{MC}}, \\
A_{3}= & (3\sqrt{3}/2)\gamma^{2}_{g}r^{2}_{\textmd{MC}},
\end{split}
\end{equation}
where $A_i$ denotes the area of the region $\texttt{A}_i$.

From~\eqref{eq_12}, a simple approximate formula for the instantaneous SIR is obtained by replacing the instantaneous total interference power in the denominator of~\eqref{eq_3} by the average total interference power. The following SIR bounds are the obtained as a function of distance $r_{0}$:
\begin{equation} \label{eq_13}
\frac{\sigma^{2}_{M}h_{0}(r_{0})}{\hat{I}_{\,u}^{\,(avg)}(r_{0})}\lesssim
\mathtt{SIR}(r_{0})\lesssim\frac{\sigma^{2}_{M}h_{0}(r_{0})}{\hat{I}_{\,l}^{\,(avg)}(r_{0})} ~.
\end{equation}

From~\eqref{eq_1}-\eqref{eq_2}, the lower and upper bounds on SIR can be rewritten as \footnote{In (11)-(13) we use $\simeq$ (as opposed to $=$) and $\lesssim$ (as opposed to $\leq$) to emphasize that the polynomials are approximations to the bounds.}
\begin{equation} \label{eq_14}
\mathtt{SIR}_{\,l/u}(r_{0})\simeq\underbrace{\left[
\frac{(\max\{r_{\textmd{ref}},r_{0}\}/r_{\textmd{MC}})^{-\alpha}}
{\exp(\sigma^{2}_{z}/2)\sum_{i=0}^{P}a^{u/\,l}_i(r_{0}/r_{\textmd{MC}})^{P-i}}
\right]}_{\xi_{\,l/u}}z\,,
\end{equation}
where $z$ is a log-normal random variable with $z\sim \mathcal{LN}(\mu_{z}=0,\sigma_{z}=(0.1\ln10)\sigma_{L})$. We note that
%
%
the SIR expression in~\eqref{eq_14} can be written as a function of the normalized distance $\bar{r}_{0}=r_{0}/r_{\textmd{MC}}$ as well. Thus, for a given $\bar{r}_{0}$, the lower and upper bounds on SIR are modelled as log-normal random variables,
\begin{equation} \label{eq_15}
\mathtt{SIR}_{\,l/u}(\bar{r}_{0})\sim \LN(\mu_{\,\rm{SIR}_{\,l/u}}(\bar{r}_{0}),\sigma_{\,\rm{SIR}_{\,l/u}}) \,,
\end{equation}
with $\sigma_{\,\rm{SIR}_{\,l/u}}=\sigma_{z}$, and $\mu_{\,\rm{SIR}_{\,l/u}}(\bar{r}_{0})=\ln \xi_{\,l/u}(\bar{r}_{0})$.

Assuming isolated SCs, the average number of SCs that is required to cover the total MC area in outage is given as
\begin{equation} \label{eq_16}
\begin{split}
N & =\left\lceil (A^{\rm{outage}}_{\,1}+A^{\rm{outage}}_{\,2}+A^{\rm{outage}}_{\,3})/A_{\textmd{SC}} \right\rceil \\
&=\left\lceil (\delta^{\,(1)} A_{\,1}+\delta^{\,(2)} A_{\,2}+\delta^{\,(3)} A_{\,3})/A_{\textmd{SC}} \right\rceil
\end{split}
\end{equation}
where $A^{\rm{outage}}_{\,i}$ is the average area within the region $\texttt{A}_{i}$ that is in outage and $\delta^{(i)} = A^{\rm{outage}}_{\,i}/A_i$ is the fractional outage area within region $\texttt{A}_{i}$. Finally, $\lceil x \rceil$ denotes the smallest integer greater than or equal to $x$.
Using the identities for $A_{i},i=1,2,3$ in~\eqref{eq_4}, the average number of SCs required is
\begin{equation} \label{eq_17}
\begin{split}
N =\left\lceil \left( \delta^{\,(1)} (1-\gamma_{g})^{2} + 2\delta^{\,(2)}\gamma_{g}(1-\gamma_{g}) + \delta^{\,(3)}\gamma_{g}^{2}\right) \frac{r^{2}_{\textmd{MC}}}{r^{2}_{\textmd{SC}}} \right\rceil
\end{split}
\end{equation}
in which we assume that $A_{\textmd{SC}}$ is an hexagon with circumscribed radius $r_{\textmd{SC}}$.

We now obtain the fractional area in outage associated with the regions $\texttt{A}_{i};i=1,2,3$, i.e., $\delta^{(1)}$, $\delta^{(2)}$, and $\delta^{(3)}$. As in the previous section, $\delta^{(i)},\,i=1,2,3$ can be calculated using the triangular wedge within the MC where $0\leq \theta_0 \leq \pi/6$ (the triangle ADF in Fig.~\ref{Fig_2_new}(b)). Now, the triangle ABH and the rectangle BDEH in Fig.~\ref{Fig_2_new}(b) correspond to $(1/12)$th of the area of $\texttt{A}_{1}$, and $\texttt{A}_{2}$, respectively. Similarly, the triangle HEF corresponds to $(1/12)$th of the area of $\texttt{A}_{3}$.

\subsection{Outages within region $\texttt{A}_1$}

Any user within the triangle ABH defining $\texttt{A}_{1}$ is associated with the BS at the center of the cell $\textmd{BS}_{0}$. We, therefore, only need to concern ourselves with the signal power and received SIR, denoted as $\textmd{SIR}_{0}$, from this BS. From~\eqref{eq_7} and~\eqref{eq_15}, the upper/lower bounds on the ROP at a point at normalized distance $\bar{r}_{0}$  ($=r_0/r_\textmd{MC}$, where $r_0$ is the distance from $\textmd{BS}_0$), is
\begin{equation} \label{eq_18}
\begin{split}
& \Px[C_{\,l/u}<C_{0}]^{\textmd{ABH}} (\bar{r}_{0}) =\Px[\mathtt{SIR}_{\,l/u}<\Gamma(2^{C_{0}}-1)] \\
& = \! \Phi \! \left( \! \frac{\ln(\Gamma(2^{C_{0}}-1))-\mu_{\,\mathtt{SIR}_{\,l/u}}(\bar{r}_{0})}
{\sigma_{\,\mathtt{SIR}_{\,l/u}}} \! \right) \\
& = \! \Phi \! \left(\! \frac{\ln(\Gamma(2^{C_{0}}-1))-\ln \xi_{\,l/u}(\bar{r}_{0})}{\sigma_{z}}\!\right) \, \\
& = \! \Phi \! \left( \! \frac{1}{\sigma_{z}}\! \left[\ln(\Gamma(2^{C_{0}}-1)) \!- \! \ln \frac{\max\{\bar{r}_{\textmd{ref}},\bar{r}_{0}\}^{-\alpha}}
{\exp(\sigma^{2}_{z}/2)\sum_{i=0}^{P}a^{\,u/\,l}_i\bar{r}_{0}^{P-i}} \! \right] \! \right)
 \end{split}
\end{equation}
where $\Phi(y)=1-Q(y)$ is the cumulative distribution function (CDF) of the standard normal distribution and $\bar{r}_{\textmd{ref}}=r_{\textmd{ref}}/r_{\textmd{MC}}$.

A point in $\texttt{A}_{1}$ is in outage if the ROP at that point is greater than threshold $\eta$, i.e., $\Px[C<C_{0}]^{\textmd{ABH}}_{\,l/u}(\bar{r}_{0})>\eta$. The ROP in~\eqref{eq_18} is an increasing function of $\bar{r}_{0}$. Thus, in order to obtain the area of a MC which is in outage, it is enough to obtain the value of $\bar{r}_{0}$ that gives $\Px[C<C_{0}]^{\textmd{ABH}}_{\,l/u}(\bar{r}^{\,\rm{opt}}_{0})=\eta$. Any point outside this distance would be in outage. Since $\Px[C<C_{0}]^{\textmd{ABH}}_{\,l/u}(\bar{r}_{0})$ is not an easily invertible function in $\bar{r}_{0}$, the threshold value $\bar{r}^{\rm{opt}}_{0,l/u}$ can be obtained numerically, by solving the following equation for $\bar{r}_{0}$
\begin{equation} \label{eq_19}
\begin{split}
& \Phi\left(\frac{1}{\sigma_{z}}\left[\ln(\Gamma(2^{C_{0}}-1))-\ln \frac{\max\{\bar{r}_{\textmd{ref}},\bar{r}_{0}\}^{-\alpha}}
{\exp(\sigma^{2}_{z}/2)\sum_{i=0}^{P}a^{u/\,l}_{i}\bar{r}_{0}^{P-i}}\right]\right) \\
& =\eta.
\end{split}
\end{equation}
It is worth noting that it is sufficient to obtain the optimal value of $\bar{r}^{\,\rm{opt}}_{0,l/u}$ only once and store the resulting values for any choices of $C_{0}$ and $\sigma_{L}$ in a look-up-table.

\begin{figure*}[t]
 \begin{eqnarray}
 \nonumber \\  \hline  \nonumber \\
 \delta^{\,(1)}_{\,l/u}  &=& \frac{\texttt{A}^{\rm{outage}}_{1}}{\texttt{A}_{1}} = 1- \frac{2\,(r^{\,\rm{opt}}_{0,\,l/u})^{2}}{\sqrt{3}\,r^{2}_{\texttt{A}_{1}}} \left\{
 \begin{array}{l c}
 \frac{\pi}{3} & 0\leq \frac{r^{\,\rm{opt}}_{0,\,l/u}}{r_{\texttt{A}_{1}}} \leq \frac{\sqrt{3}}{2} \\
 \frac{\pi}{3}-2[\cos^{-1}\frac{\sqrt{3}\,r_{\texttt{A}_{1}}}
 {2\,r^{\,\rm{opt}}_{0,\,l/u}}-\frac{1}{2}\sin (2\cos^{-1}\frac{\sqrt{3}\,r_{\texttt{A}_{1}}}{2\,r^{\,\rm{opt}}_{0,\,l/u}})] & \frac{\sqrt{3}}{2}\leq \frac{r^{\,\rm{opt}}_{0,\,l/u}}{r_{\texttt{A}_{1}}} \leq 1 \\
 0 & \text{otherwise}
 \end{array}
 \right. \!.
 \label{eq_20} \\
 \delta^{\,(1)}_{\,l/u} & = & 1- \frac{2\,(\bar{r}^{\,\rm{opt}}_{0,\,l/u})^{2}}{\sqrt{3}\,(1-\gamma_{g})^{2}} \left\{
 \begin{array}{l c}
 \frac{\pi}{3} & 0\leq \frac{\bar{r}^{\,\rm{opt}}_{0,\,l/u}}{(1-\gamma_{g})} \leq \frac{\sqrt{3}}{2} \\
 \frac{\pi}{3}-2[\cos^{-1}\frac{\sqrt{3}\,(1-\gamma_{g})}
 {2\,\bar{r}^{\,\rm{opt}}_{0,\,l/u}}-\frac{1}{2}\sin (2\cos^{-1}\frac{\sqrt{3}\,(1-\gamma_{g})}{2\,\bar{r}^{\,\rm{opt}}_{0,\,l/u}})] & \frac{\sqrt{3}}{2}\leq \frac{\bar{r}^{\,\rm{opt}}_{0,\,l/u}}{(1-\gamma_{g})} \leq 1 \\
 0 & \text{otherwise}
 \end{array}
 \right. \,.
 \label{eq_21} \\
 \hline \nonumber
 \end{eqnarray}
 \end{figure*}

Now, if $r^{\textmd{opt}}_{0,\,l/u}=\bar{r}^{\textmd{opt}}_{0,\,l/u}r_{\textmd{MC}}$ is less that  $\frac{\sqrt{3}}{2}r_{\texttt{A}_{1}}$, the circle $x_{0}^{2}+y_{0}^{2}=(r^{\,\rm{opt}}_{0,\,l/u})^{2}$ is entirely inside the hexagonal area $\texttt{A}_{1}$. As a result, $A^{\rm{outage}}_{1}=A_{1}-\pi(r^{\,\rm{opt}}_{0,\,l/u})^{2}$. On the other hand, if $\frac{\sqrt{3}}{2}r_{\texttt{A}_{1}} \leq r^{\,\rm{opt}}_{0,\,l/u} \leq r_{\texttt{A}_{1}}$, the circle $x_{0}^{2}+y_{0}^{2}=(r^{\,\rm{opt}}_{0,\,l/u})^{2}$ is cut off by the edges of the hexagonal region $\texttt{A}_{1}$. The area that is cut of by edges is given by
\[A_{\rm{cut}} = 6(r^{\,\rm{opt}}_{0,\,l/u})^{2} [\cos^{-1}\frac{\sqrt{3}\,r_{\texttt{A}_{1}}} {2\,r^{\,\rm{opt}}_{0,\,l/u}}-\frac{1}{2}\sin (2\cos^{-1}\frac{\sqrt{3}\,r_{\texttt{A}_{1}}}{2\,r^{\,\rm{opt}}_{0,\,l/u}})].\]
Therefore, the outage area is obtained from \[A^{\rm{outage}}_{1}=A_{1}-A^{\rm{coverage}}_{1} = A_{1}-\pi(r^{\,\rm{opt}}_{0,\,l/u})^{2}+A_{\rm{cut}},\]
and the fractional area of $\texttt{A}_{1}$ in outage, $\delta^{(1)}_{l/u}$, is given by Eq.~\eqref{eq_20} (appeared at the top of the next page).

Using $r_{\texttt{A}_{1}}=(1-\gamma_{g})r_{\textmd{MC}}$ and $r^{\textmd{opt}}_{0,\,l/u}=\bar{r}^{\textmd{opt}}_{0,\,l/u}r_{\textmd{MC}}$, $\delta^{(1)}_{l/u}$ can now be written as Eq.~\eqref{eq_21} (appeared at the top of the next page). The expression in~\eqref{eq_21} indicates that in a reuse-1 interference-limited network, the bounds on the fractional outage area within region $\texttt{A}_1$, $\delta^{\,(1)}_{\,l/u}$, do not depend on the MC radius or the BS transmit power.

\subsection{Outages within region $\texttt{A}_2$}

A user within the rectangle BDEH in Fig.~\ref{Fig_2_new} receives service from the stronger of $\textmd{BS}_{0}$ and $\textmd{BS}_{1}$. The bounds on the ROP are given by
\begin{equation} \label{eq_22}
\begin{split}
& \Px[C_{l/u}<C_{0}]^{\textmd{BDEH}} =\Px[\max\limits_{j=0,1} \,\mathtt{SIR}_{\,\textit{j}\,{l/u}}<\Gamma(2^{C_{0}}-1)] \\
& =\Px[\mathtt{SIR}_{\,0\,(l/u)}<\Gamma(2^{C_{0}}-1)]\,
\Px[\mathtt{SIR}_{\,1\,(l/u)}<\Gamma(2^{C_{0}}-1)] \\
& = \prod_{j=0}^{1} \Phi\left(\frac{1}{\sigma_{z}}\left[\rho-\ln \frac{\bar{r}_{j}^{(-\alpha)}}
{\exp(\sigma^{2}_{z}/2)\sum_{i=0}^{P}a^{\,u/\,l}_i\bar{r}_{j}^{P-i}}\right]\right)
\end{split}
\end{equation}
\begin{figure*}[ht]
 \begin{eqnarray}
 \nonumber \\  \hline  \nonumber \\
\delta^{\,(2)}_{\,l/u} & \approx & \int_{0}^{b} \int_{-g}^{0}\Px[C_{l/u}<C_{0}]^{\textmd{BDEH}}f_{\rm{BDEH}}(x,y)\textmd{d}x \textmd{d}y \nonumber \\
& =  & \frac{1}{gb} \int_{0}^{b} \int_{-g}^{0} \left[ \Phi\left(\frac{1}{\sigma_{z}}\left[\rho-\ln \frac{[\left(((\sqrt{3}/2)r_{\textmd{MC}}-x)^{2}+y^{2}\right)^{0.5}/r_{\textmd{MC}}]^{-\alpha}}
{\exp(\sigma^{2}_{z}/2)\sum_{i=0}^{P}a^{u/\,l}_i
[\left(((\sqrt{3}/2)r_{\textmd{MC}}-x)^{2}+y^{2}\right)^{0.5}/r_{\textmd{MC}}]^{P-i}}\right]\right) \right. \label{eq_23}  \\
& & \hspace{5em} \left. \times \Phi\left(\frac{1}{\sigma_{z}}\left[\rho-\ln \frac{[\left(((\sqrt{3}/2)r_{\textmd{MC}}+x)^{2}+y^{2}\right)^{0.5}/r_{\textmd{MC}}]^{-\alpha}}
{\exp(\sigma^{2}_{z}/2)\sum_{i=0}^{P}a^{u/\,l}_i
    [\left(((\sqrt{3}/2)r_{\textmd{MC}}+x)^{2}+y^{2}\right)^{0.5}/r_{\textmd{MC}}]^{P-i}}\right]\right) \right] \textmd{d}x \textmd{d}y \nonumber \\
    \nonumber \\
\delta^{\,(2)}_{\,l/u} & \approx & B \int_{0}^{\frac{1-\gamma_{g}}{2}} \int_{-\frac{\sqrt{3}\gamma_{g}}{2}}^{0} \left[ \Phi\left(\frac{1}{\sigma_{z}}\left[\rho-\ln \frac{[\left(((\sqrt{3}/2)-\bar{x})^{2}+\bar{y}^{2}\right)^{0.5}]^{-\alpha}}
{\exp(\sigma^{2}_{z}/2)\sum_{i=0}^{P}a^{u/\,l}_i
[\left(((\sqrt{3}/2)-\bar{x})^{2}+\bar{y}^{2}\right)^{0.5}]^{P-i}}\right]\right) \right. \label{eq_24} \\
& & \hspace{7em} \left. \times \Phi\left(\frac{1}{\sigma_{z}}\left[\rho-\ln \frac{[\left(((\sqrt{3}/2)+\bar{x})^{2}+\bar{y}^{2}\right)^{0.5}]^{-\alpha}}
{\exp(\sigma^{2}_{z}/2)\sum_{i=0}^{P}a^{u/\,l}_i
[\left(((\sqrt{3}/2)+\bar{x})^{2}+\bar{y}^{2}\right)^{0.5}]^{P-i}}\right]\right) \right] \textmd{d}\bar{x} \textmd{d}\bar{y} \nonumber \\
 \hline \nonumber
 \end{eqnarray}
 \end{figure*}
where $\rho = \ln[\Gamma(2^{C_{0}}-1)]$. In~\eqref{eq_22}, $\bar{r}_{0}$ and $\bar{r}_{1}$ are the associated normalized distances to $\textmd{BS}_{0}$ and $\textmd{BS}_{1}$, respectively. The points within BDEH have nearly equal distances to $\textmd{BS}_{0}$ and $\textmd{BS}_{1}$, so they have approximately equal ROPs and we approximate $\delta^{\,(2)}_{\,l/u}$ with the average ROP within BDEH. To obtain the average ROP within BDEH, it is more convenient to represent the distances $r_{0}$ and $r_{1}$ in Cartesian coordinates. With the origin set at the middle of the segment connecting $\textmd{BS}_{0}$ and $\textmd{BS}_{1}$, the average of $\Px[C_{l/u}<C_{0}]^{\textmd{BDEH}}$ is given by Eq.~\eqref{eq_23} (appeared in the next page), where $f_{\rm{BDEH}}(x,y) = 1/gb$ is the, assumed uniform, distribution of users within BDEH. With the change of variables $\bar{x}=x/r_{\textmd{MC}}$ and $\bar{y}=y/r_{\textmd{MC}}$, $\delta^{\,(2)}_{\,l/u}$ can be obtained equivalently from Eq.~\eqref{eq_24} (appeared in the next page), where $B=\frac{4}{3\sqrt{3}\gamma_{g}(1-\gamma_{g})}$. This integral can be
%
evaluated numerically.
As with the region $\texttt{A}_1$, from~\eqref{eq_24}, $\delta^{\,(2)}_{\,l/u}$ (the fractional area in outage in region $\texttt{A}_2$) is independent of the MC radius and the BS transmit power.

\subsection{Outages within region $\texttt{A}_3$}

The final region to analyze is $\texttt{A}_3$ defined by the triangle HEF. A user in $\texttt{A}_3$ receives service from the strongest of $\textmd{BS}_{0}$, $\textmd{BS}_{1}$ and $\textmd{BS}_{2}$. Therefore,
\begin{equation} \label{eq_25}
\begin{split}
& \Px[C_{l/u}<C_{0}]^{\textmd{HEF}}  = \Px[\max\limits_{j=0,1,2} \,\mathtt{SIR}_{\,j\,(l/u)}<\Gamma(2^{C_{0}}-1)] \\
& = \prod_{j=0}^{2} \Px[\mathtt{SIR}_{\,j\,(l/u)}<\Gamma(2^{C_{0}}-1)]  \\
& = \prod_{j=0}^{2} \underbrace{\Phi\left(\frac{1}{\sigma_{z}}\left[\rho -\ln \frac{\bar{r}_{j}^{(-\alpha)}}{\exp(\sigma^{2}_{z}/2)
\sum_{i=0}^{P}a^{u/\,l}_i\bar{r}_{j}^{P-i}}\right]\right)}_{\Phi_{j}}
\end{split}
\end{equation}
where $r_{2}=\left(x^{2}+(1.5\,r_{\textmd{MC}}-y)^{2}\right)^{0.5}$ is the associated distance to $\textmd{BS}_{2}$. As in the case for $\texttt{A}_{2}$, the average ROP within HEF well approximates $\delta^{\,(3)}_{\,l/u}$ in $\texttt{A}_{3}$. Thus, $\delta^{\,(3)}_{\,l/u}$ is approximated as
\begin{equation} \label{eq_26}
\begin{split}
\delta^{\,(3)}_{\,l/u} & \approx \int_{-g}^{0} \int_{b}^{\frac{\sqrt{3}}{3}(x+g)+b} \Px[C_{l/u}<C_{0}]^{\textmd{HEF}}f_{\textmd{HEF}}(x,y)\textmd{d}y \textmd{d}x \\
& = \frac{1}{A_{\textmd{HEF}}} \int_{-g}^{0} \int_{b}^{\frac{\sqrt{3}}{3}(x+g)+b}
\Phi_{0}\,\Phi_{1}\,\Phi_{2} ~\textmd{d}y \,\textmd{d}x,
\end{split}
\end{equation}
where $A_{\textmd{HEF}}=\sqrt{3}\gamma^{2}_{g}r^{2}_{\textmd{MC}}/8$ is the area of triangle HEF. Since the terms $\Phi_{0}$,$\Phi_{1}$, and $\Phi_{2}$ do not depend on $r_{\textmd{MC}}$, with the change of variables $\bar{x}=x/r_{\textmd{MC}}$ and $\bar{y}=y/r_{\textmd{MC}}$, the dependence of $\delta^{\,(3)}_{\,l/u}$ on the MC radius can be seen easily via
\begin{equation} \label{eq_27}
\delta^{\,(3)}_{\,l/u} \approx \frac{8}{\sqrt{3}\gamma^{2}_{g}} \int_{-\frac{\sqrt{3}\gamma_{g}}{2}}^{0} \int_{\frac{1-\gamma_{g}}{2}}^{\frac{\sqrt{3}}{3}(\bar{x}+
\frac{\sqrt{3}\gamma_{g}}{2})+\frac{1-\gamma_{g}}{2}}
\Phi_{0}\,\Phi_{1}\,\Phi_{2} ~\textmd{d}\bar{y} \,\textmd{d}\bar{x},
\end{equation}
Again, the above integration can be evaluated numerically.

Finally, having evaluated $\delta^{\,(1)}_{\,l/u}$, $\delta^{\,(2)}_{\,l/u}$, $\delta^{\,(3)}_{\,l/u}$ at hand, the lower/upper bound on the fractional area of the MC in outage is obtained from
\begin{equation} \label{eq_30}
\begin{split}
\delta^{\textmd{MC}}_{l/u}= \delta^{\,(1)}_{\,l/u} (1-\gamma_{g})^{2} + 2\delta^{\,(2)}_{\,l/u}\gamma_{g}(1-\gamma_{g}) +\delta^{\,(3)}_{\,l/u}\gamma^{2}_{g}
\end{split}
\end{equation}

From~\eqref{eq_21},~\eqref{eq_24},~\eqref{eq_27}-\eqref{eq_30} we see that in a reuse-1 \emph{interference-limited} network, the percentage area in outage (without SCs in use) does not depend on the MC radius and BS transmit power, but does depend on the path loss exponent $\alpha$, variance of log-normal fading $\sigma^{2}_{L}$, rate threshold $C_{0}$, and ROP threshold $\eta$. Our simulations will show that the approximations used in our derivation are valid when $r_{\textmd{MC}} \lesssim 2\,\textmd{km}$.

We emphasize that our analysis does not address the \emph{  specific placement of SCs} within a MC. The analysis here averages over the shadowing, i.e., we obtain the average area that would be in outage. A specific placement is possible for \textit{a given realization of the log-normal shadowing}. With a hex-tile arrangement of SCs within each MC, for a given realization of log-normal shadowing, if the achieved rate from the serving BS falls below the rate threshold $C_0$ with probability greater than $\eta$ (averaged over the small-scale fading), then a SC is placed at that location. Moreover, the analysis here assumes \textit{isolated} SCs, i.e., SCs that do not interfere with the BSs or with each other. In practice, SCs generally share frequencies (frequency reuse-1 amongst SCs) though they may be co-channel or orthogonal with the BSs. In either case, there is some interference within the SC layer and a residual outage area; in the next section this area is evaluated via simulations.

\section{Numerical Results}
\label{sec:Simulations}

This section presents the results of simulations that validate the approximations used in the previous section, illustrate the results in terms of average number of SCs required and the residual outage area. However, we begin with illustrating the impact of the model for BS locations in the network. Specifically, we cover the PPP and hexagonal lattice models and compare the coverage probability to that from real-world deployments.

\subsection{Comparison of Outage Probability under Different BS Location Models}

\begin{figure}[ht]
 \begin{center}
 \subfigure[Rate outage probability]{\includegraphics[width = .45\textwidth]{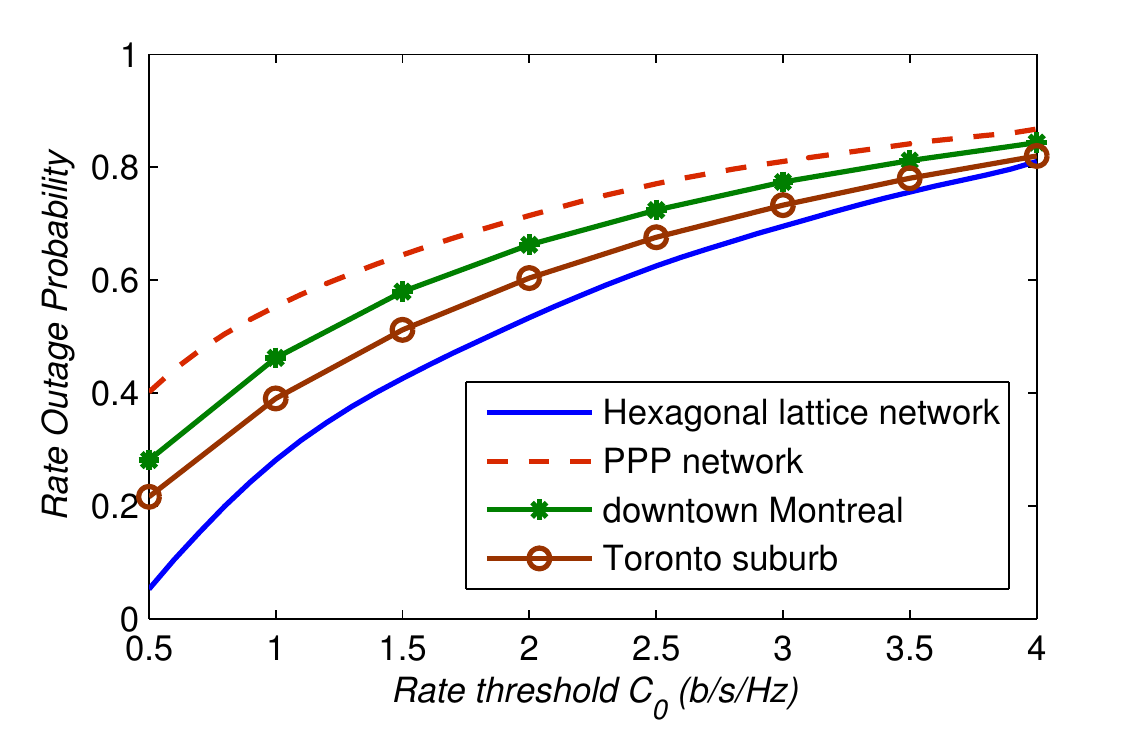}}
 \subfigure[Required number of SCs]{\includegraphics[width = .45\textwidth]{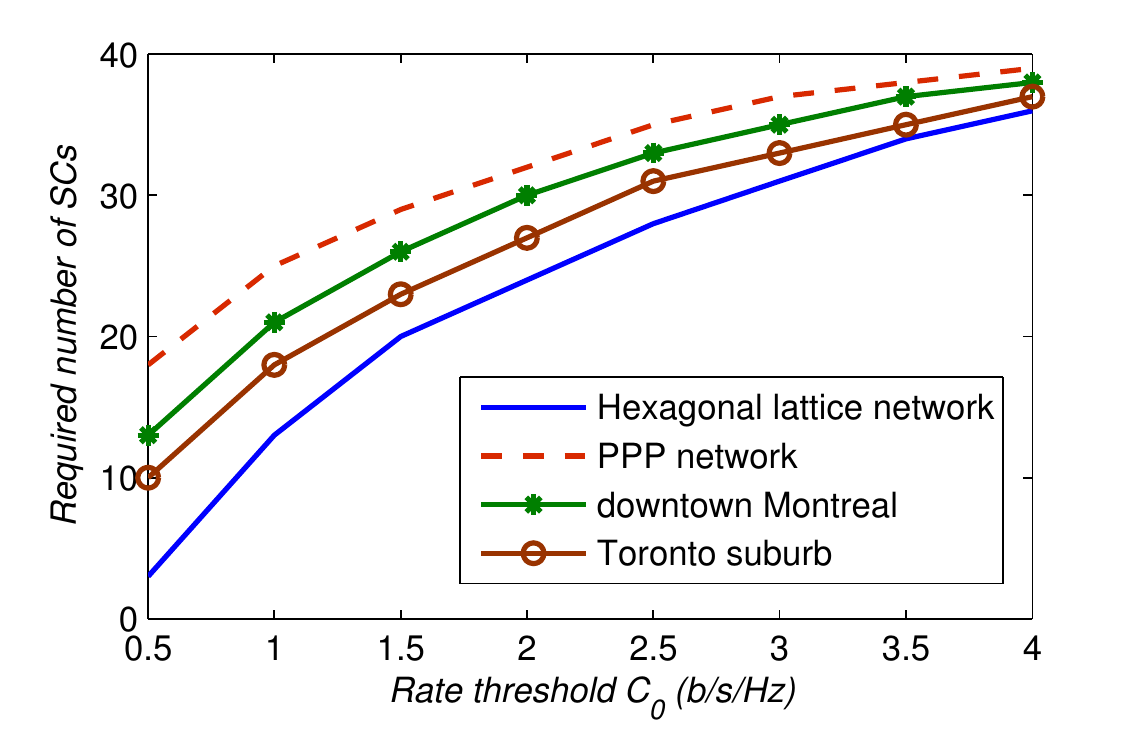}}
 \caption{Comparison of the: (a) Rate outage probability; (b) Required number of SCs within the example area of $S=2.59\times 10^{6} ~\textmd{m}^2$, in hexagonal lattice network, PPP network, and two real-world BS deployments from downtown Montreal and Toronto suburb. }
\label{fig_PPP}
\end{center}
\end{figure}

As mentioned in the introduction, when compared to real-world BS deployments, the hexagonal lattice and PPP networks provide lower and upper bounds on the outage probability in the network, respectively~\cite{guo2013spatial,andrews2011tractable}. Correspondingly, these two models would, respectively, provide a lower and an upper bound on the required number of SCs needed in cellular networks. As discussed in the introduction, an analysis of dependent placements of SCs in multi-layer PPP networks is not available in the literature. Therefore we cannot make a fair comparison between our two-layer hexagonal-lattice network with the two-layer dependent PPP network presented in~\cite{Ref11}. However, with no SCs in use, we can compare a hexagonal-lattice network with a PPP network based on their resulting average outage probability (averaged over the network area). Figure~\ref{fig_PPP}-(a) compares these lower and upper bounds on the outage probability with that of two real-world BS deployments (the associated data for the locations of BSs is gathered from a publicly available Canadian database www.ic.gc.ca/spectrumdirect). The examples are from a suburb of Toronto and from downtown Montreal, in Canada. Table~\ref{Tab_data points details} provides the details of these point sets.
\begin{table*}
\caption{Details of the points sets representing the location of BSs}
\centering
\begin{tabular}{|c||c|c|c|c|c|}
   \hline
    $\textmd{Point Set}$  &  $\textmd{Operator}$ & $\textmd{Frequency\,(MHz)}$ & $\textmd{Center Location (latitude, longitude)}$ & $\textmd{Area} \,(\textmd{m}\times \textmd{m})$ & $\textmd{BS density} \,(\#\,\textmd{BSs}/\textmd{m}^2)$ \\
 \hline \hline
   $\textmd{Toronto suburb}$ & $\textmd{Bell}$ & $2630$  & $43.6671^{\circ} \textmd{N}, -79.5836^{\circ} \textmd{W} $ &        $6000 \times 6000$ &        $ 6.95 \times 10^{-7}$  \\  \hline
   $\textmd{downtown Montreal}$  &  $\textmd{Telus}$  & $1740$ & $45.5235^{\circ} \textmd{N}, -73.6010^{\circ} \textmd{W} $ &        $8000 \times 5500$ &          $ 1.18 \times 10^{-6} $ \\  \hline
 \end{tabular}
 \label{Tab_data points details}
 \end{table*}

To make the comparison, we assume that the two given real-world BS deployments follow the reuse-1 and isotropic radiation policy used in the analysis with $\alpha =4$ and $\sigma_L= 4~\textmd{dB}$. We obtain the corresponding average outage probability for the PPP and the two given point sets via simulation (a theoretical analysis of outage with a PPP model is only available when ignoring shadowing). As is seen from Fig.~\ref{fig_PPP}-(a), the PPP model provides a better estimate of the empirical performance in downtown Montreal which is more densely deployed than a suburb in Toronto. This result is in line with the finding reported earlier in~\cite{lu2015stochastic} that the PPP approach becomes more accurate as the network becomes denser.

The associated required number of SCs for each of the above considered cases is illustrated in Fig.~\ref{fig_PPP}-(b). To calculate these numbers, we consider the example area of $S = 2.59\times 10^{6} ~\textmd{m}^2$ (corresponding to the area of an hexagon with outer radius $r_{\textmd{MC}}=1000 ~\rm{m}$), obtain the total outage area (by multiplying the area $S$ with the average outage probability) and then divide the associated total outage area by the area of a single SC (with $r_{\textmd{SC}}=150 ~\rm{m}$).

The results shown in Fig.~\ref{fig_PPP} are important in understanding the motivation for this paper - while assuming a PPP for BS locations simplifies analysis and simulations, what remains unknown is how tight the bound is. Now, given the analysis in this paper from the hexagonal lattice model for the locations of BSs, the number of required SCs (albeit isolated SCs) in real-world deployments is bounded from below and above. Further, the figures emphasize that neither the PPP nor the hexagonal models are any better than the other - they each provide a different bound.

We now subsection the efficacy of our analysis. The parameters common to the simulations are presented in Table~\ref{tab:sims}.

\begin{table}[h]
\caption{Parameters Used in the Simulations}
 \label{tab:sims}
 \vspace*{-0.2in}
 \begin{center}
 \begin{tabular}{|c|c||c|c|}
   \hline
   Parameter & Value & Parameter & Value\\
 \hline \hline
 MC transmit power & 43 dBm & SC transmit power & 20 dBm \\ \hline
 Additive noise power & -100 dBm & SNR-gap ($\Gamma$) & 2 dB \\ \hline
 ROP threshold ($\eta$) & 0.5 &  $\sigma_L$ & 4 dB  \\ \hline
 $\alpha$ & 3, 3.8, 4 & $\gamma_g$ & 0.25 \\ \hline
 \end{tabular}
 \end{center}
 \end{table}

\subsection{Validating the Approximations and Tightness of Bounds}

\begin{figure}[t]
 \begin{center}
 \subfigure[the effect of rate threshold]{\includegraphics[width = .45\textwidth]{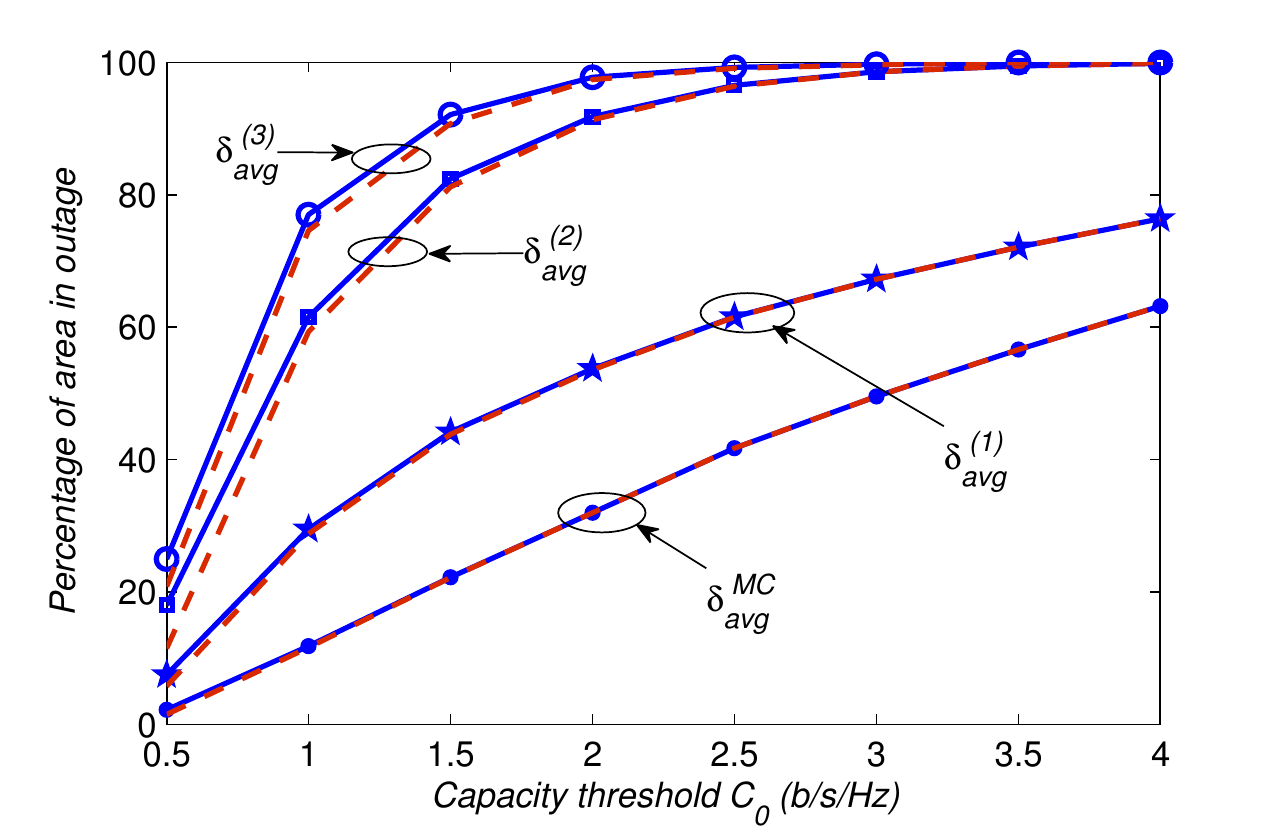}}
 \subfigure[the effect of MC radius]{\includegraphics[width = .45\textwidth]{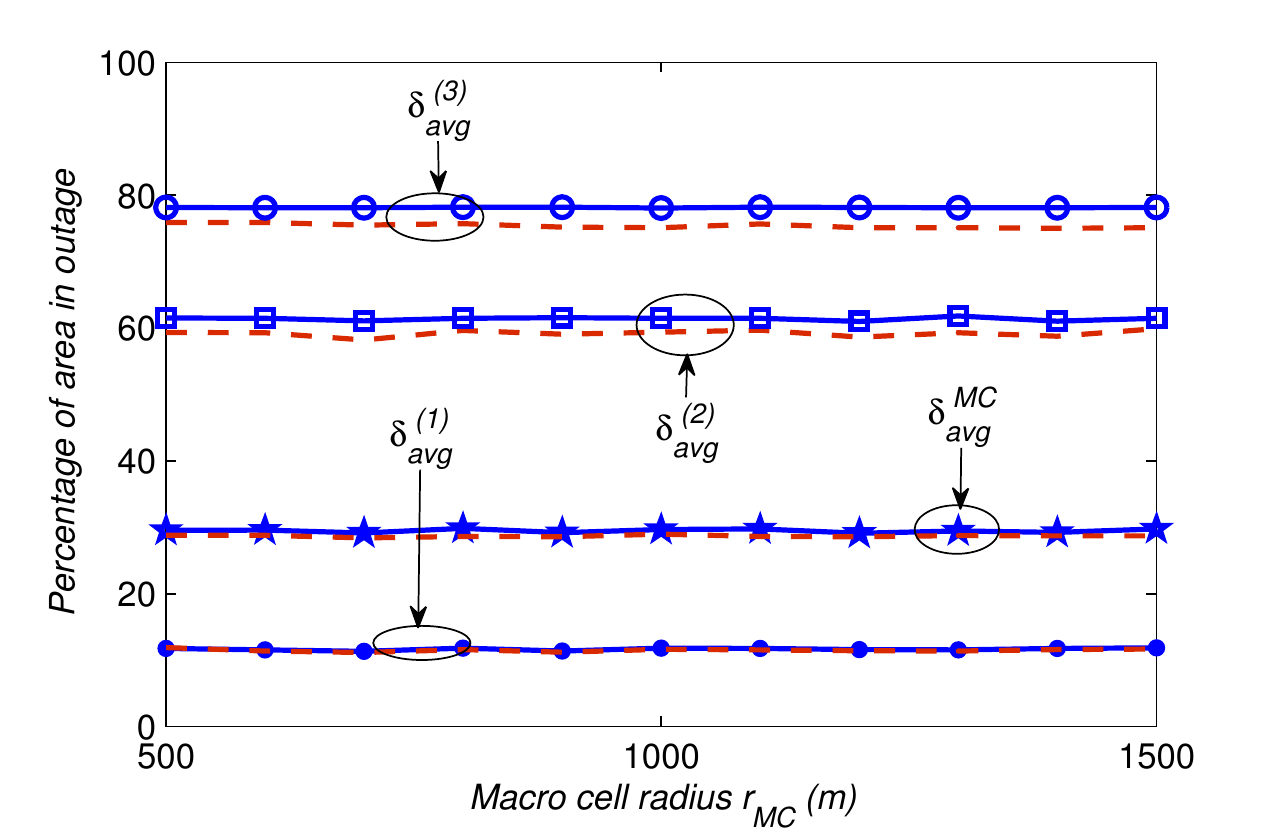}}
 \caption{Comparison of the analytical results (solid lines) with the exact results obtained from Monte Carlo simulations (dotted lines) for different values of: (a) spectral efficiency threshold $C_{0}$; (b) MC radius $r_{\textmd{MC}}$ ($C_{0}=1 ~\textmd{b/s/Hz}$).}
 \label{new_fig_5}
 \end{center}
 \end{figure}

We first evaluate the accuracy of the presented analytical formulation. Fig.~\ref{new_fig_5} plots the fractional area in outage using the analytic results from Section~\ref{sec:NoOfSCs} (the solid lines in the figure), and compares these results with the exact results obtained from Monte Carlo simulations (the dotted lines in the figure), for a PLE of $\alpha=4$. The figure plots the fractional area in outage within each region of Fig.~\ref{Fig_2_new} and the overall fraction in outage. The analytical results are the average of the lower and upper bounds, indicated by the subscript 'avg' in the figure. As is clear from Fig.~\ref{new_fig_5}, there is a close match between the approximate analytic and simulated results for all cases with the differences approaching zero for moderate to high values of desired spectral efficiency $C_{0}$. In particular, for the example of $C_{0}=1 ~\textmd{b/s/Hz}$, the relative error in $\delta^{\textmd{MC}}_{\textmd{avg}}$ is 2.5\%. Therefore, a  design based on the closed-form expressions provided in Sections~\ref{sec:NoOfSCs} is valid. Fig.~\ref{new_fig_5}-(b) confirms that the fractional area in outage is independent of the macrocell radius.

\begin{figure}[t]
\begin{center}
\includegraphics[width=0.5\textwidth]{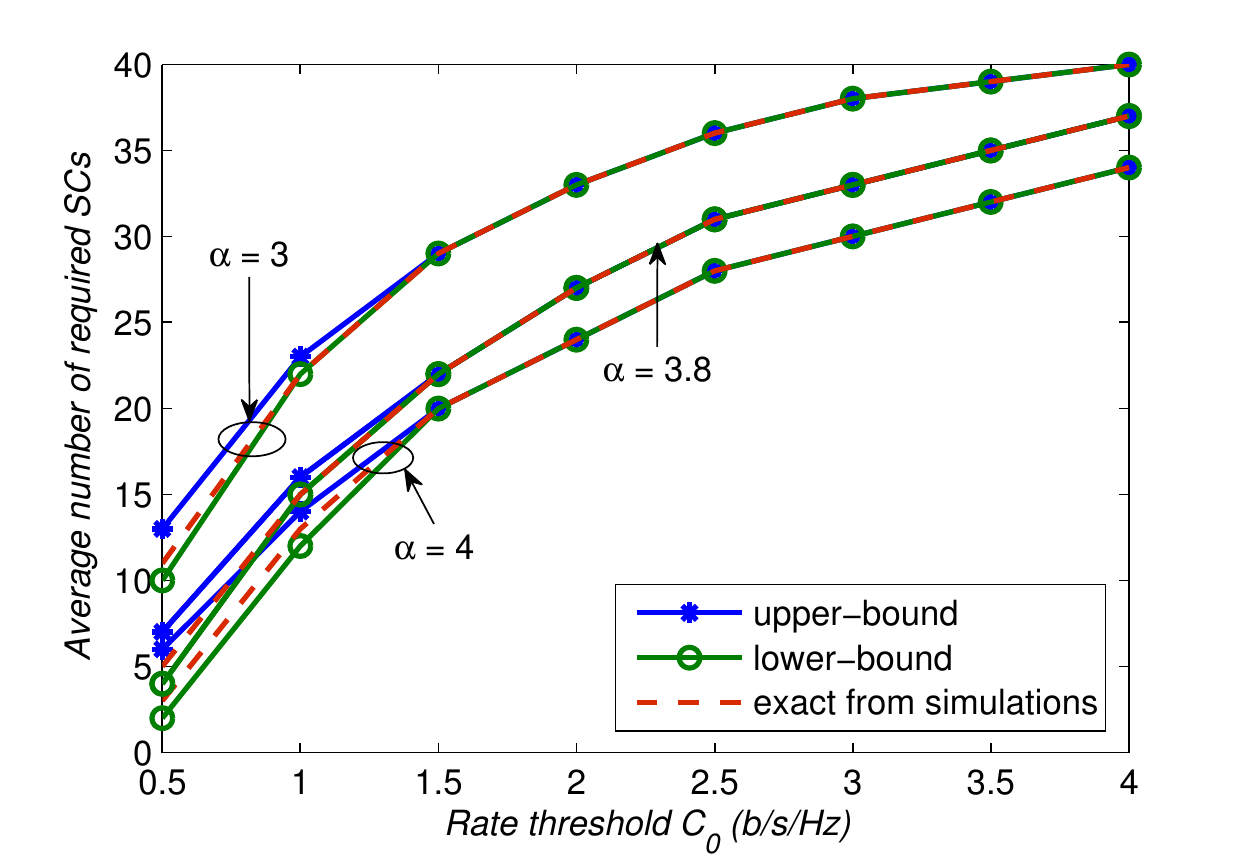}
\caption{Lower and upper bounds on the required number of SCs for different values of rate threshold $C_{0}$ in a reuse-1 network with $r_{\textmd{MC}}=1 ~\rm{km}$, and $r_{\textmd{SC}}=150 ~\rm{m}$. The exact results from simulations using the instantaneous interference and SIR from~\eqref{eq_3} is also included for comparison in dashed lines.}
\label{new_Fig_6}
\end{center}
\end{figure}

To illustrate the tightness of the bounds and the efficacy of the analysis in predicting the average number of required SCs, Fig.~\ref{new_Fig_6} plots the lower and upper bounds obtained in Section~\ref{sec:NoOfSCs} as a function of the spectral efficiency threshold $C_{0}$, with $r_{\textmd{MC}}=1 ~\rm{km}$ and SC radius $r_{\textmd{SC}}=150 ~\rm{m}$. The exact results from simulations using the instantaneous interference (and instantaneous SIR) from~\eqref{eq_3} are included as dashed lines. For $C_{0}\gtrsim 1.5 ~\rm{b/s/Hz}$, the lower and upper bounds are indistinguishable from the simulations. For smaller values of $C_{0}$, the bounds are tight and the average of the lower and upper bounds is a close approximation to the required number of SCs. We note that since the required number of SCs is directly related to the fractional area in outage (shown in~\eqref{eq_17}), we conclude that the average of the lower and upper bounds on the fractional area in outage, given in~\eqref{eq_30}, accurately approximates the fractional area in outage obtained from simulations. Therefore, in in the rest of the figures, we take the average of the provided lower and upper bounds as the result for a hexagonal-lattice network.


\subsection{Design Examples for the Placement of SCs}

\begin{figure}[t]
\begin{center}
\includegraphics[width=0.45\textwidth]{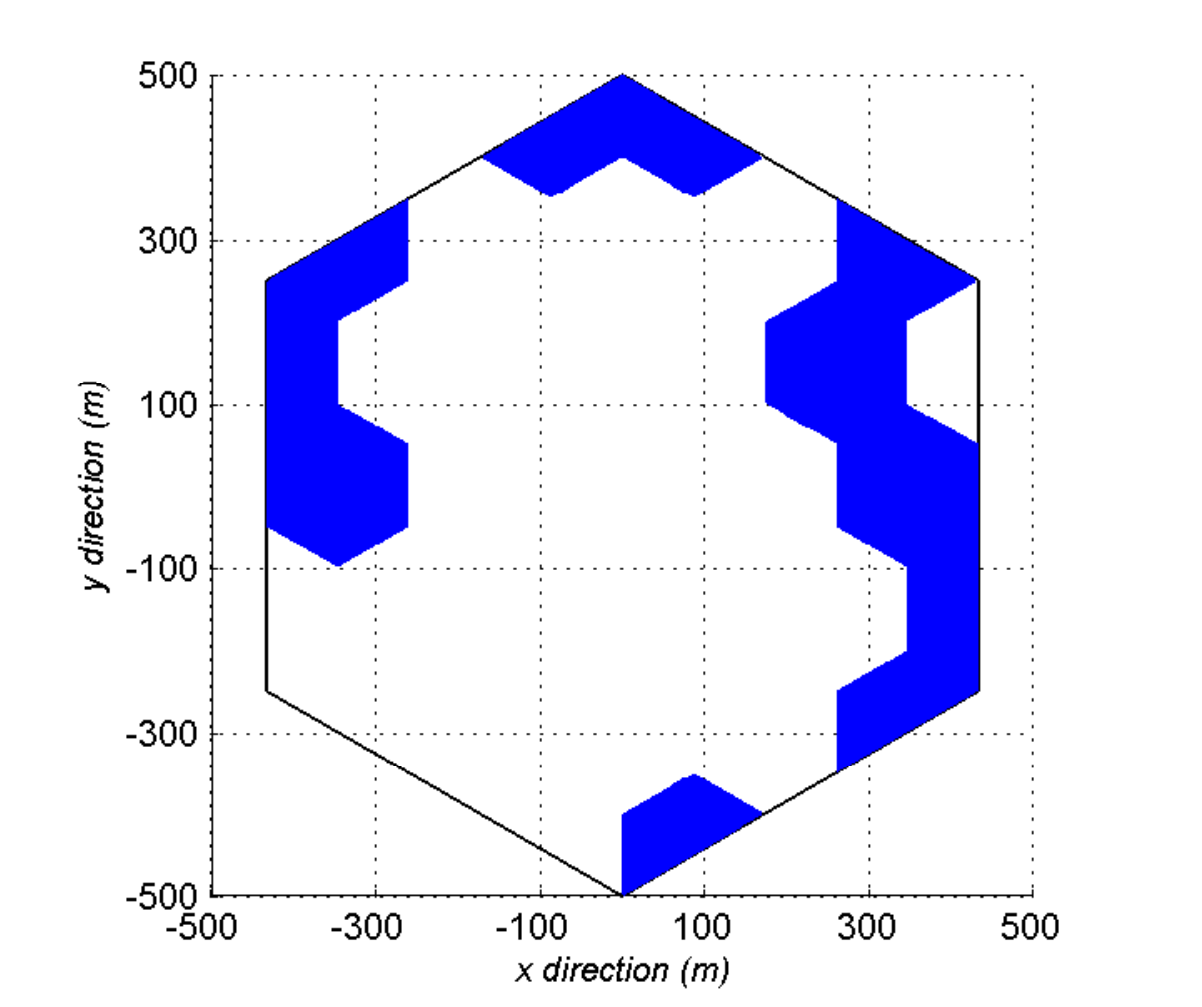}
\caption{The hexagonal candidate areas for SC placement to cover outages within a MC under a typical realization of log-normal shadowing. The example is for $C_0= 1~\rm{b/s/Hz}$, $r_{\textmd{MC}}=500 ~\rm{m}$, and $r_{\textmd{SC}}=150 ~\rm{m}$.}
\label{new_Fig_7}
\end{center}
\end{figure}

Having illustrated the accuracy of our analysis, we now show its use in a system design and the residual outage. Figure~\ref{new_Fig_7} illustrates the outage area within a MC for \emph{one realization} of log-normal shadowing for the target spectral efficiency of $C_0= 1~\rm{b/s/Hz}$ in a reuse-1 network with $r_{\textmd{MC}}=500 ~\rm{m}$, and $r_{\textmd{SC}}=100 ~\rm{m}$. Here, the percentage of MC area in outage without SCs in use, is 29\%. Now that the outage area is identified, the associated SCs are placed accordingly. In this example, as counted via Fig.~\ref{new_Fig_7}, the number of required isolated SCs to completely cover the outage area is 7. In this count, the SCs at the edges shared between two MCs, are counted as half. This number of SCs is close to the averaged results,  given by Section~\ref{sec:NoOfSCs} as $N_{\textmd{avg}} = 8$.

Users within SCs are covered by SC APs, while the other users connect to the BS. If isolated SCs are used, the outage area is covered completely by SCs and there would no residual outages in the network. However, without an interference-aware resource allocation amongst SCs, the SCs interfere with each other and the network experiences a residual outage area. The residual outage area depends on whether co-channel or orthogonal (with respect to BS) SCs are used.

\begin{figure}[t]
 \begin{center}
 \subfigure[co-channel SCs]{\includegraphics[width = .47\textwidth]{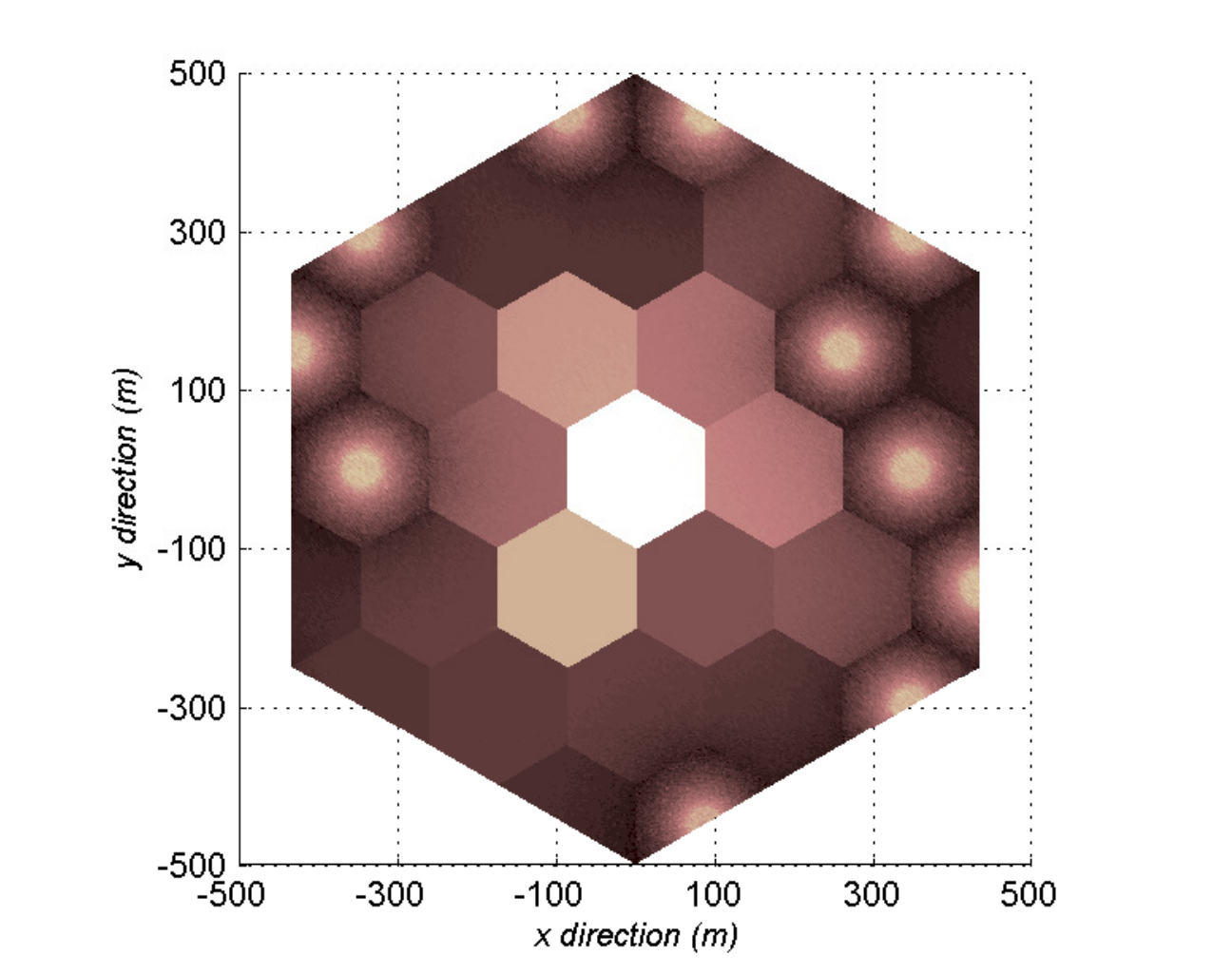}}
 \subfigure[orthogonal SCs]{\includegraphics[width = .47\textwidth]{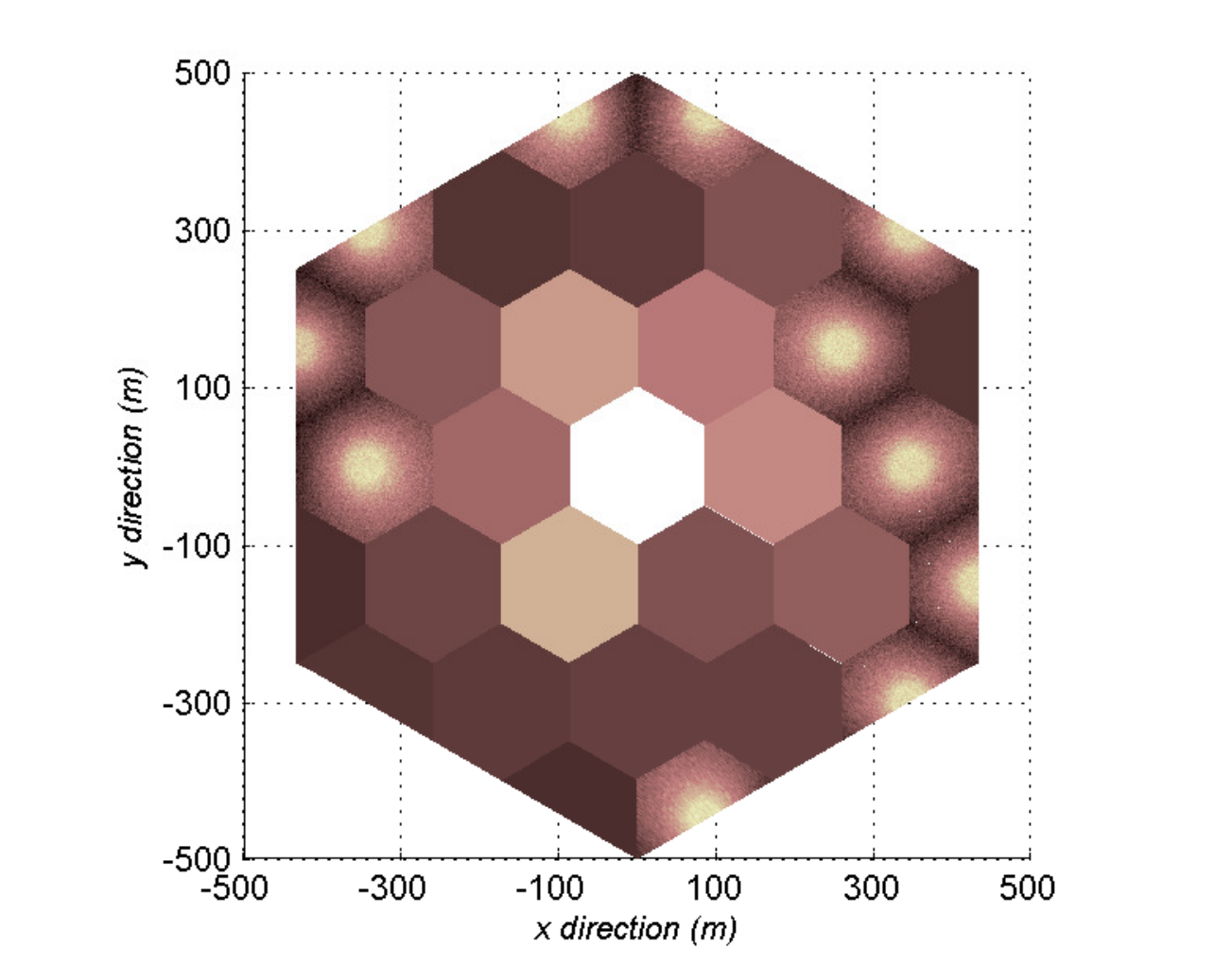}}
 \caption{Ergodic rate within a MC in reuse-1 network. The target spectral efficiency is $C_0= 1~\rm{b/s/Hz}$, the MC radius is $r_{\textmd{MC}}=500 ~\rm{m}$, and $r_{\textmd{SC}}=150 ~\rm{m}$.}
 \label{new_Fig_8}
 \end{center}
 \end{figure}
\begin{figure}[ht]
 \begin{center}
 \subfigure[co-channel SCs: 26\% residual outage area]{\includegraphics[width = .47\textwidth]{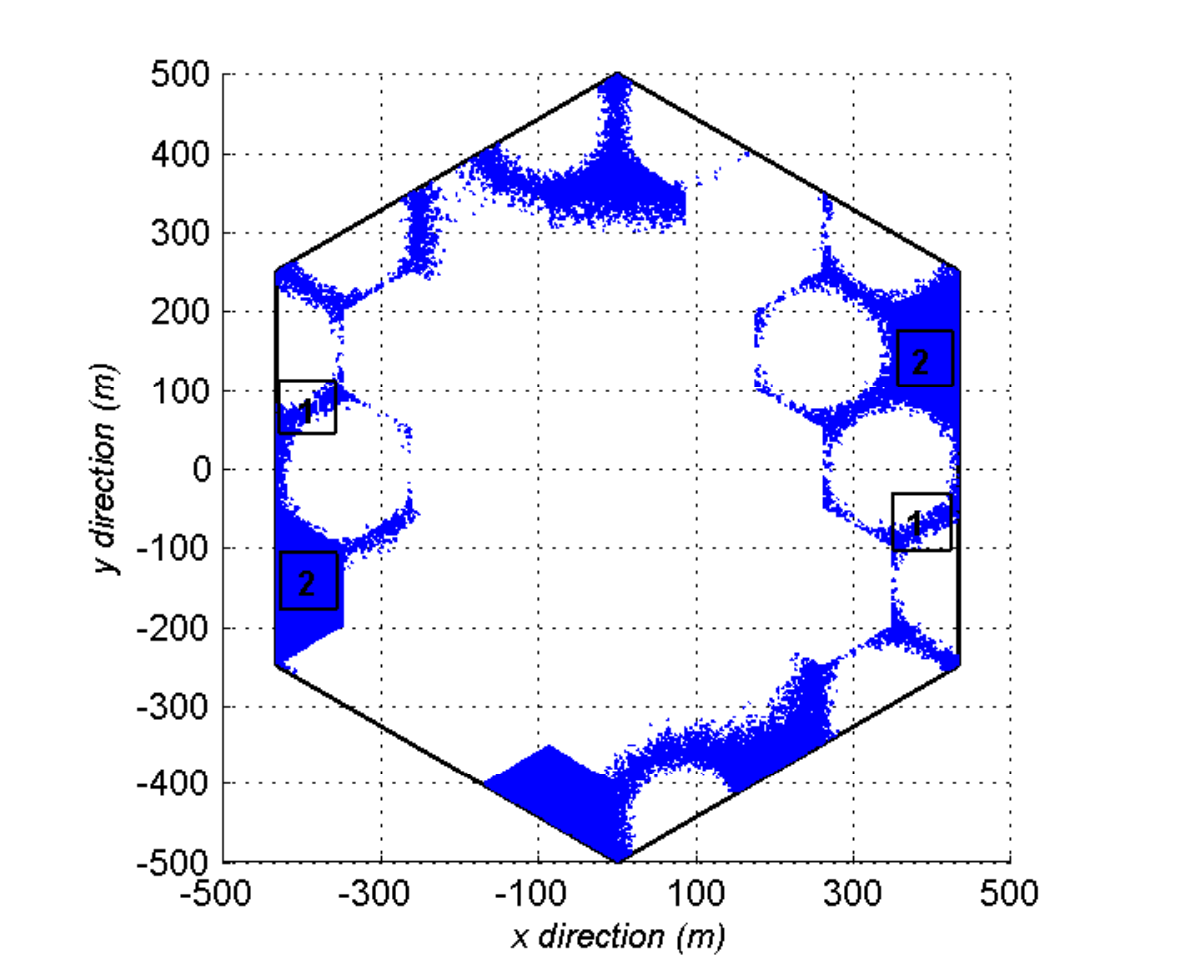}}
 \subfigure[orthogonal SCs: 4\% residual outage area]{\includegraphics[width = .47\textwidth]{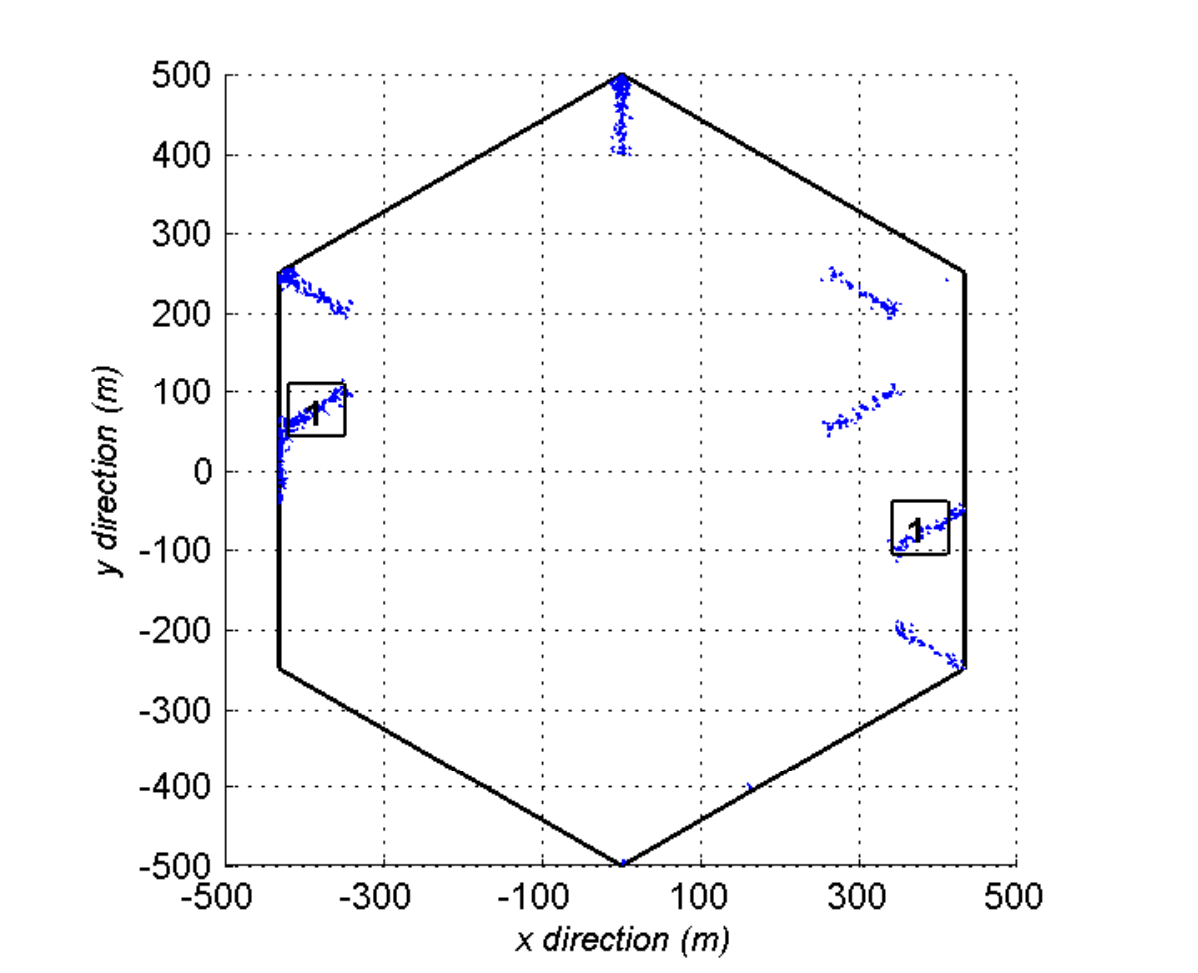}}
 \caption{The residual outage area with SCs in effect within a MC in reuse-1 network. The target spectral efficiency is $C_0= 1~\rm{b/s/Hz}$.}
 \label{new_Fig_9}
 \end{center}
 \end{figure}

Figures~\ref{new_Fig_8} and~\ref{new_Fig_9} illustrate the ergodic rate (including the effect of small-scale Rayleigh fading with unit variance) and the associated residual outage area, respectively for the same realization of log-normal shadowing as in Fig.~\ref{new_Fig_7} (obtained via simulations). Without SCs, 29\% of the MC area is in outage. Within each figure, the subplot $(a)$ plots the results with co-channel SCs and the subplot $(b)$ plots the results with orthogonal SCs.

\begin{figure}[t]
\begin{center}
\includegraphics[width=0.5\textwidth]{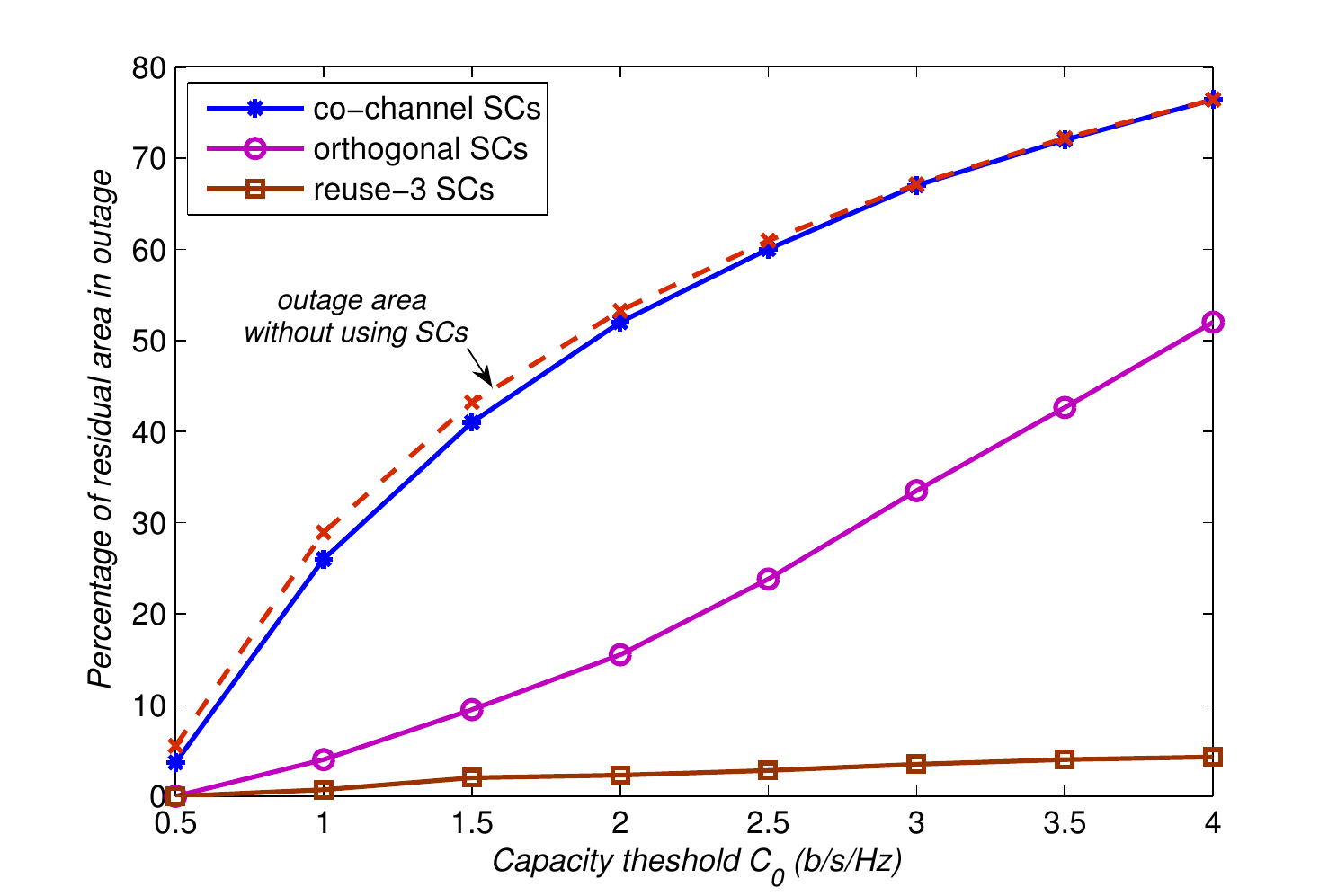}
\caption{The percentage of residual area in outage (obtained via Monte Carlo simulations) for different values of rate threshold $C_{0}$ in reuse-1 network. The percentage of area in outage without SCs is also included for comparison in dashed line.}
\label{new_Fig_10}
\end{center}
\end{figure}

With co-channel SCs, only a minor performance improvement is possible with the residual outage area falling to 26\%. With orthogonal SCs, on the other hand, only 4\% of the area remains in outage. Also, for larger values of rate threshold $C_{0}$ as seen in Fig.~\ref{new_Fig_10}, the improvement in performance diminishes with co-channel SCs. As a result, without using an interference-aware resource allocation amongst SCs, co-channel SCs do not reduce outage in the network.

The large residual outage under co-channel SCs can be explained as follows. In general, the residual outage area comprises of two regions: region (1) within SCs in which outage can be caused by co-channel interference from neighboring SCs, co-channel interference from BSs, pathloss, lognormal shadowing and small-scale fading; and region (2) where users receive a strong signal from the BS, but can be in outage either due to co-channel interference from SCs or small-scale fading. Examples of regions $(1)$ and $(2)$ are illustrated in Fig.~\ref{new_Fig_9}. With co-channel SCs, the regions of type $(2)$ are the main reason for the residual area in outage with SCs in use (as compared to outage area without SCs). In fact, regions of type $(2)$ could be in coverage from BS without SCs; however, with SCs in use, the co-channel interference from nearby SCs results in poor coverage in these regions. Essentially, using co-channel SCs causes new cell boundaries and only serves to move, but does not diminish, the outage area.

To improve coverage, one may take advantage of the use of resource allocation amongst SCs (e.g., the use of a frequency reuse factor amongst SCs). Even, an inefficient resource allocation like the use of frequency reuse-3 amongst SCs (in the traditional way as is used in the macro layer) leads to an improvement in performance in a reuse-1 (amongst the BSs) network as illustrated in Fig.~\ref{new_Fig_10} for orthogonal SCs. In this case, each SC has access to $1/3$ of the available bandwidth of $WA_{\textmd{SC}}/A_{\textmd{MC}}$, but the interference from other users within SCs is reduced.

With frequency reuse-3 amongst SCs, the residual outage area reaches 0.7\% for the target spectral efficiency of $1~\rm{b/s/Hz}$. We note that the use of frequency reuse-3 amongst SCs in the traditional way of assigning different 3 frequencies in a hexagonal-based macro layer, is an inefficient resource allocation amongst SCs. This is because some SCs may not be surrounded completely by other SCs, or there may not be any other SCs nearby (thereby allowing for a better frequency reuse factor). A more sophisticated resource allocation is beyond the scope of this paper.

\begin{figure}[t]
 \begin{center}
 \subfigure[co-channel SCs]{\includegraphics[width = .47\textwidth]{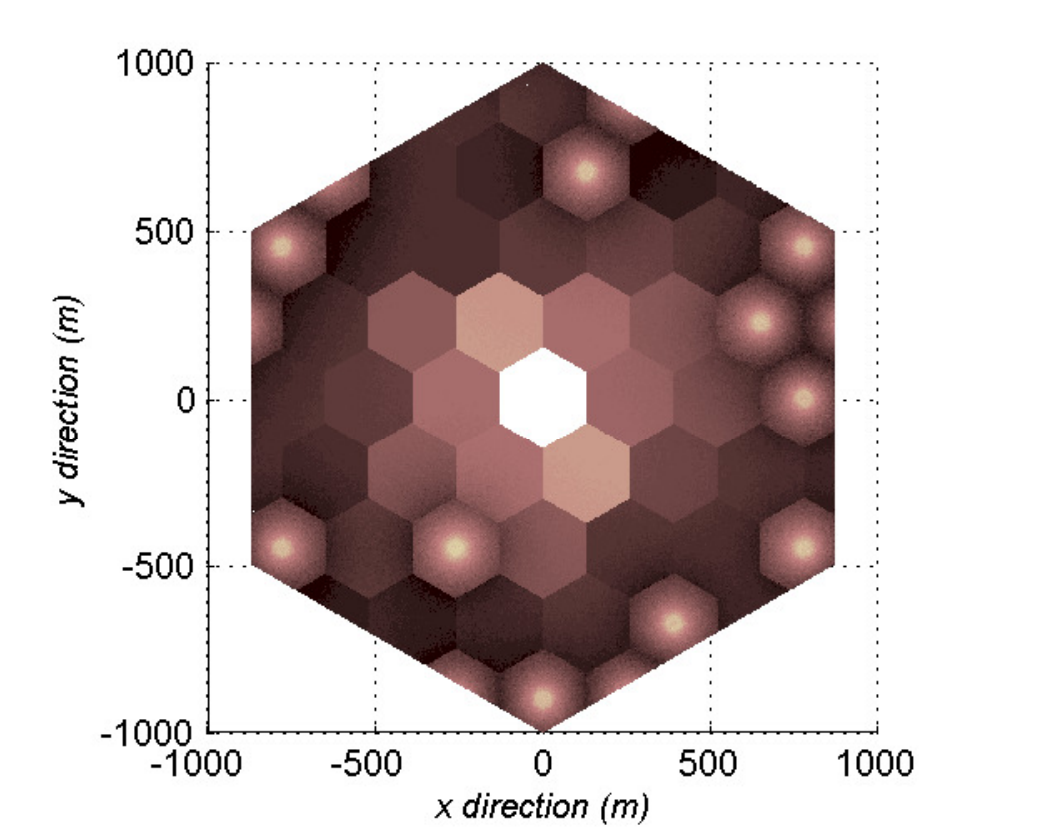}}
 \subfigure[orthogonal SCs]{\includegraphics[width = .47\textwidth]{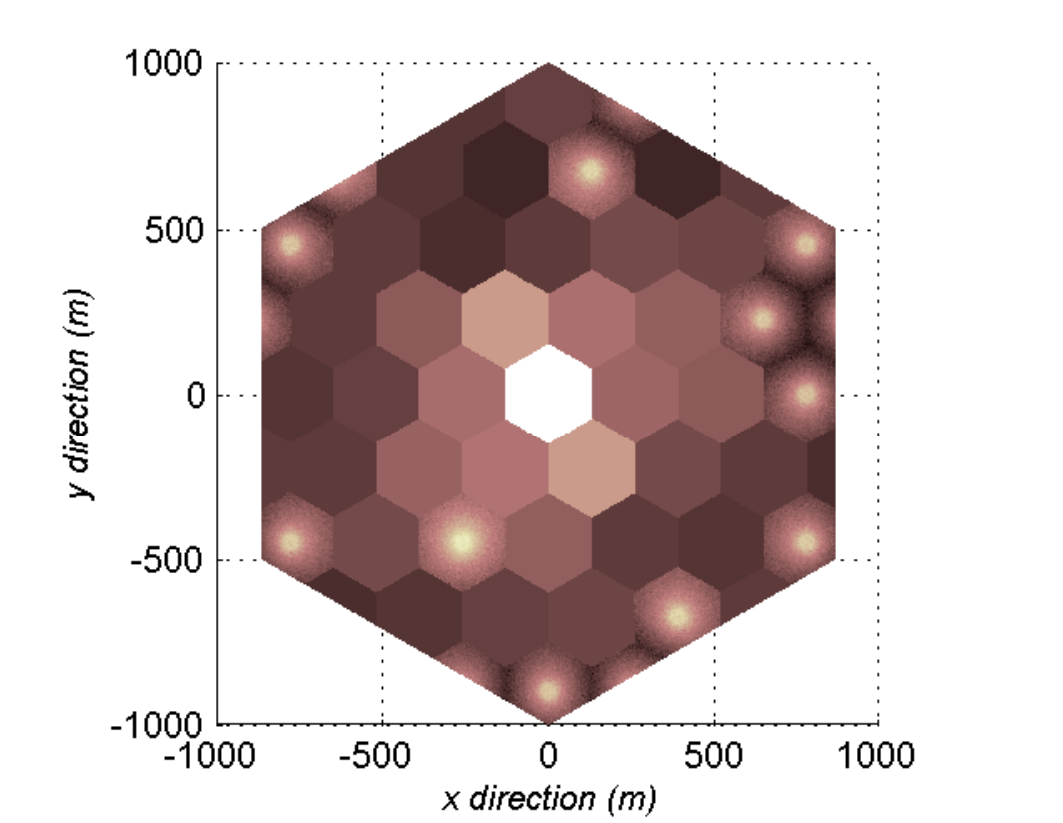}}
 \caption{Ergodic rate within a MC in reuse-1 network with $r_{\textmd{MC}}=1 ~\rm{km}$ and $r_{\textmd{SC}}=150 ~\rm{m}$. The target spectral efficiency is $C_0= 1~\rm{b/s/Hz}$.}
 \label{new_Fig_11}
 \end{center}
 \end{figure}

\begin{figure}[ht]
 \begin{center}
 \subfigure[co-channel SCs: 27\% residual outage area]{\includegraphics[width = .47\textwidth]{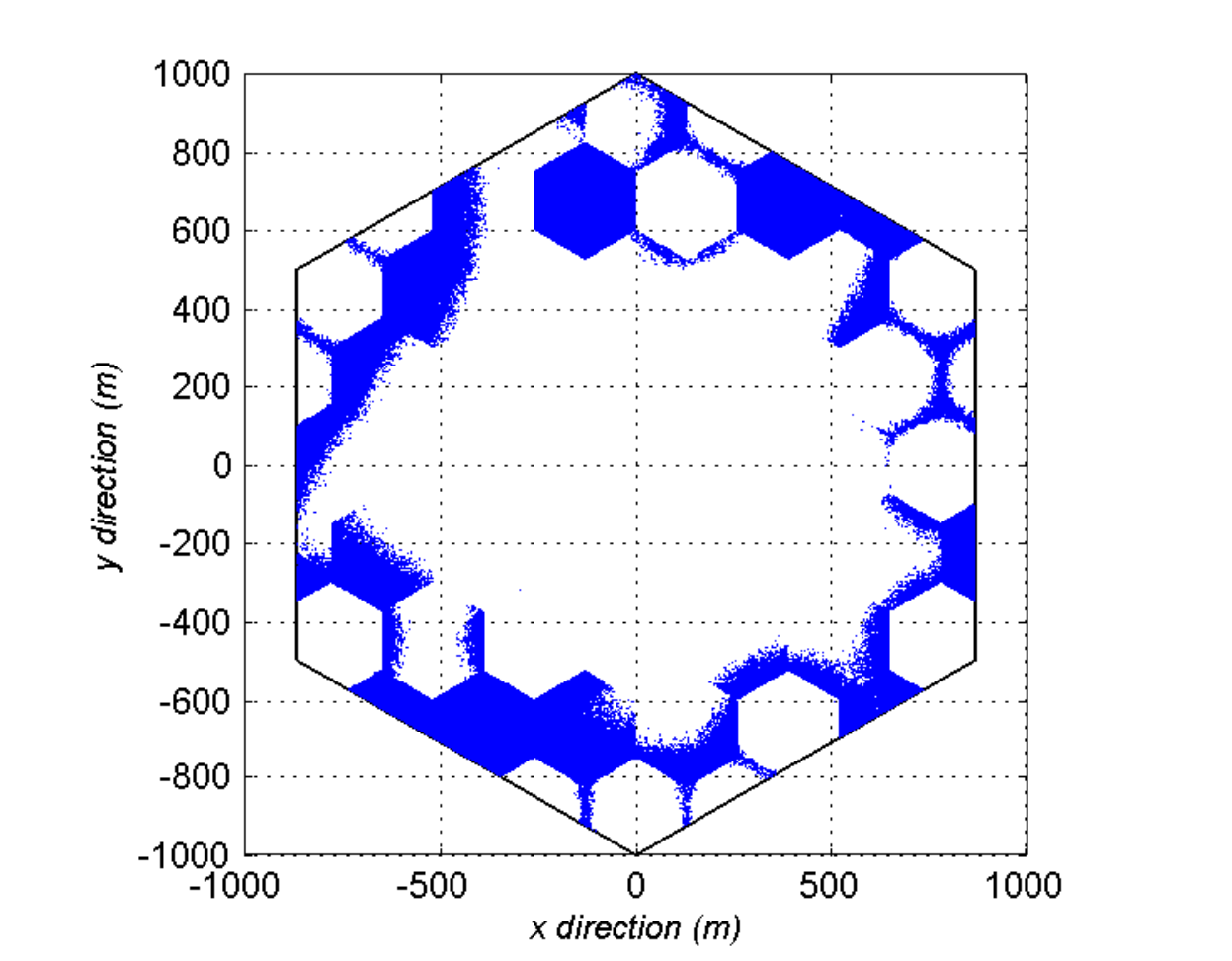}}
 \subfigure[orthogonal SCs: 4.1\% residual outage area]{\includegraphics[width = .47\textwidth]{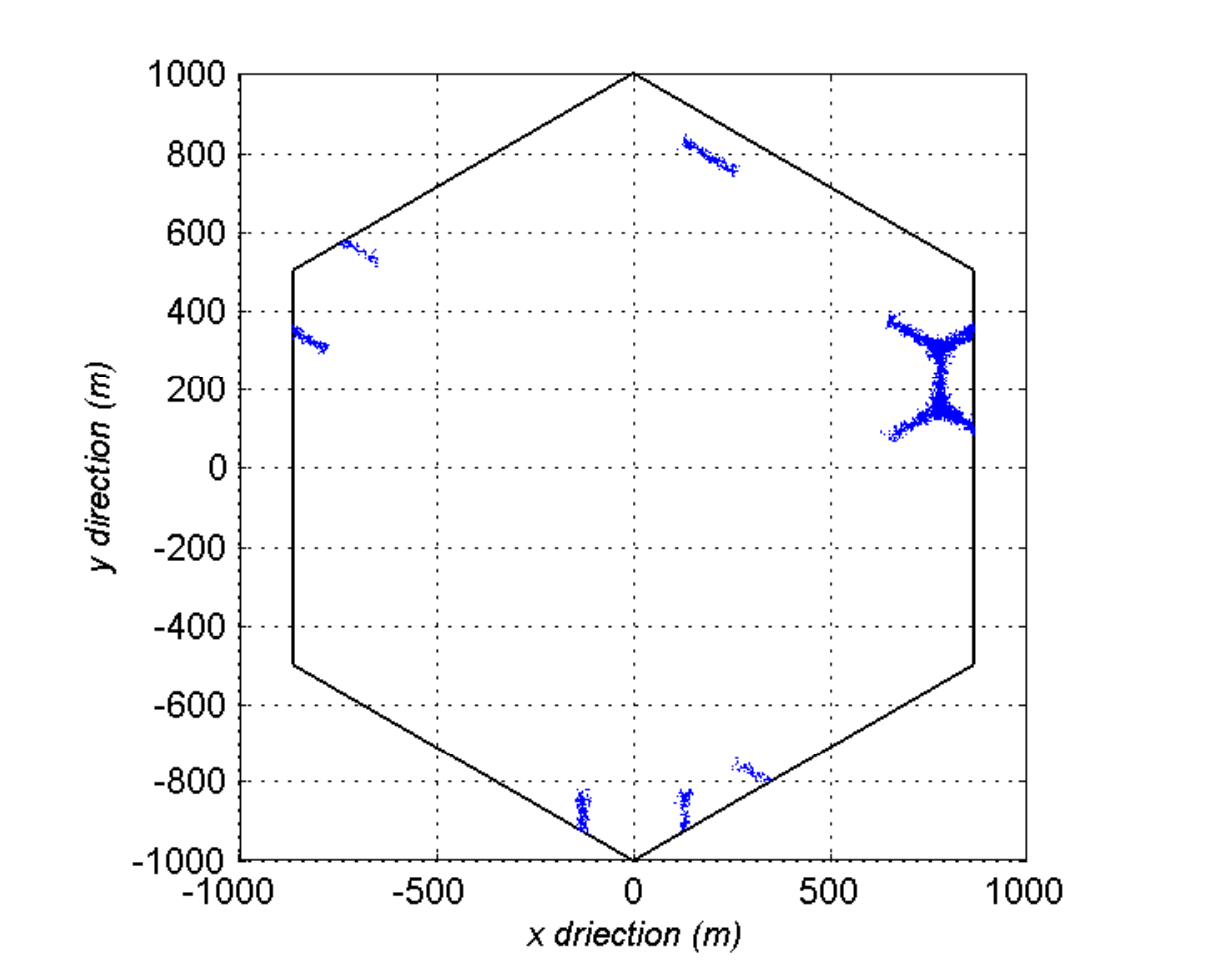}}
 \caption{The residual outage area with SCs in effect within a MC in reuse-1 network with $r_{\textmd{MC}}=1 ~\rm{km}$ and $r_{\textmd{SC}}=150 ~\rm{m}$. The target spectral efficiency is $C_0= 1~\rm{b/s/Hz}$.}
 \label{new_Fig_12}
 \end{center}
 \end{figure}

As a second design example, we consider MCs with radius of $r_{\textmd{MC}}=1 ~\rm{km}$ and SCs with $r_{\textmd{SC}}=150 ~\rm{m}$. The rest of system parameters are kept the same as in the previous example. For the target spectral efficiency of $C_{0}=1~\rm{b/s/Hz}$, the analysis (the average of lower and upper bounds in~\eqref{eq_30}) gives the percentage of MC area in outage (without SCs in use) as $\delta^{\textmd{MC}}_{\textmd{avg}}=29\%$. We note that this is the same value predicted for the percentage of outage area in the previous example. The reason is that all the network parameters, except the radius of MC, are the same as in previous example and the percentage outage area in a reuse-1 interference-limited network does not depend on the MC radius.
Now, with $r_{\textmd{MC}}=1 ~\rm{km}$ and $r_{\textmd{SC}}=150 ~\rm{m}$, Eq.~\eqref{eq_17} predicts the required number of SCs as $N_{\textmd{avg}}=\lceil \delta^{\textmd{MC}}_{\textmd{avg}}\,r^{2}_{\textmd{MC}}/r^{2}_{\textmd{SC}} \rceil = 13$. The corresponding results for the average rate and the residual outage area are illustrated in Figs.~\ref{new_Fig_11} and~\ref{new_Fig_12}. From Fig.~\ref{new_Fig_11}, the number of required (isolated) SCs is 13 which exactly matches the predicted result (though, note that the prediction is for an average).

In the latter example, the percentage of residual outage MC area is 27\% with co-channel SCs (4.1\% with orthogonal SCs). We note that the provided results are the values obtained under one realization of log-normal shadowing. The resultant values may differ for different realizations of channels. As a result the average results provided in Fig.~\ref{new_Fig_10} can be a good reference for the percentage of residual outage area for different values of target spectral efficiency under different arrangements of SCs.

\subsection{Comparison Between Reuse-1 Network and Reuse-7 Network}

In this final example we consider, via simulations, a reuse-7 BS layer. In a reuse-7 network, each MC has access only to the $1/7$ of the available bandwidth $W$, however the co-channel interference from other BSs is eliminated (considering up to two tiers of interferers). As a result, the SINR reduces to SNR. In this case, the ROP at each point of a MC is given as
\begin{equation}
\begin{split}
\Px\{C<C_{0}\}= & \Px\{(1/7)\log_{2}(1+\mathtt{SNR}_{\textmd{max}})<C_{0}\} \\
= & \Px\{\mathtt{SNR}_{\textmd{max}}<\Gamma(2^{7C_{0}}-1)\}.\
\end{split}
\end{equation}
All the expressions provided in Section IV to obtain the fractional outage area are accordingly updated  for the reuse-7 network.

\begin{figure}[t]
\begin{center}
\includegraphics[width=0.5\textwidth]{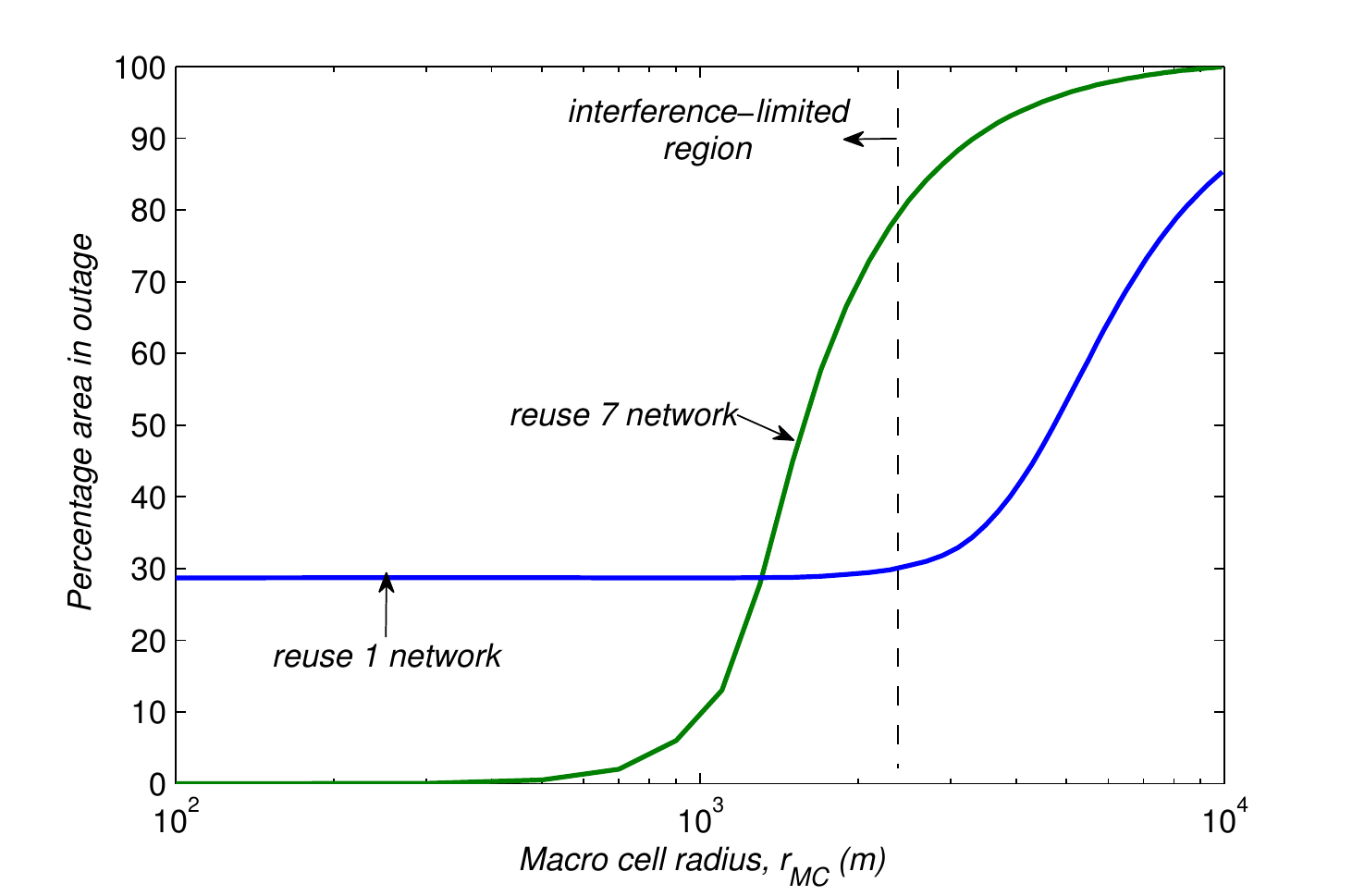}
\caption{The percentage of outage area (with no SCs in use) for different values of MC radius for the rate threshold of $C_{0}= 1 ~\rm{b/s/Hz}$ in reuse-1 and -7 networks for $\alpha = 4$}
\label{new_Fig_13}
\end{center}
\end{figure}

Figure~\ref{new_Fig_13} plots the percentage of outage area (without SCs) in reuse-1 and reuse-7 networks based on SINR for the rate threshold of $C_{0}= 1 ~\rm{b/s/Hz}$ and different values of MC radius. The region in which the reuse-1 network can be considered as interference-limited is also depicted in the figure. The results show that for large values of MC radius ($r_{\textmd{MC}}\gtrsim 2.5 \,\textmd{km}$, the reuse-1 network can not be considered as interference-limited anymore and in fact the interference power from surrounding BSs becomes comparable to the noise power $\sigma^{2}_{n}$. As a result, the percentage outage area increases with radius. However, in the interference-limited region, as expected from the formulation in Section IV, the percentage of outage area is fixed and does not depend on MC radius. As compared to a reuse-7 network, for MC radius radius $\lesssim 1$~km, a reuse-1 network sees a larger outage area. This is consistent with the increased interference with small macro-cell radii. That is, a larger number of SCs is required in a reuse-1 network to cover the area in outage as compared to a reuse-7 network.

\begin{figure}[t]
\begin{center}
\includegraphics[width=0.5\textwidth]{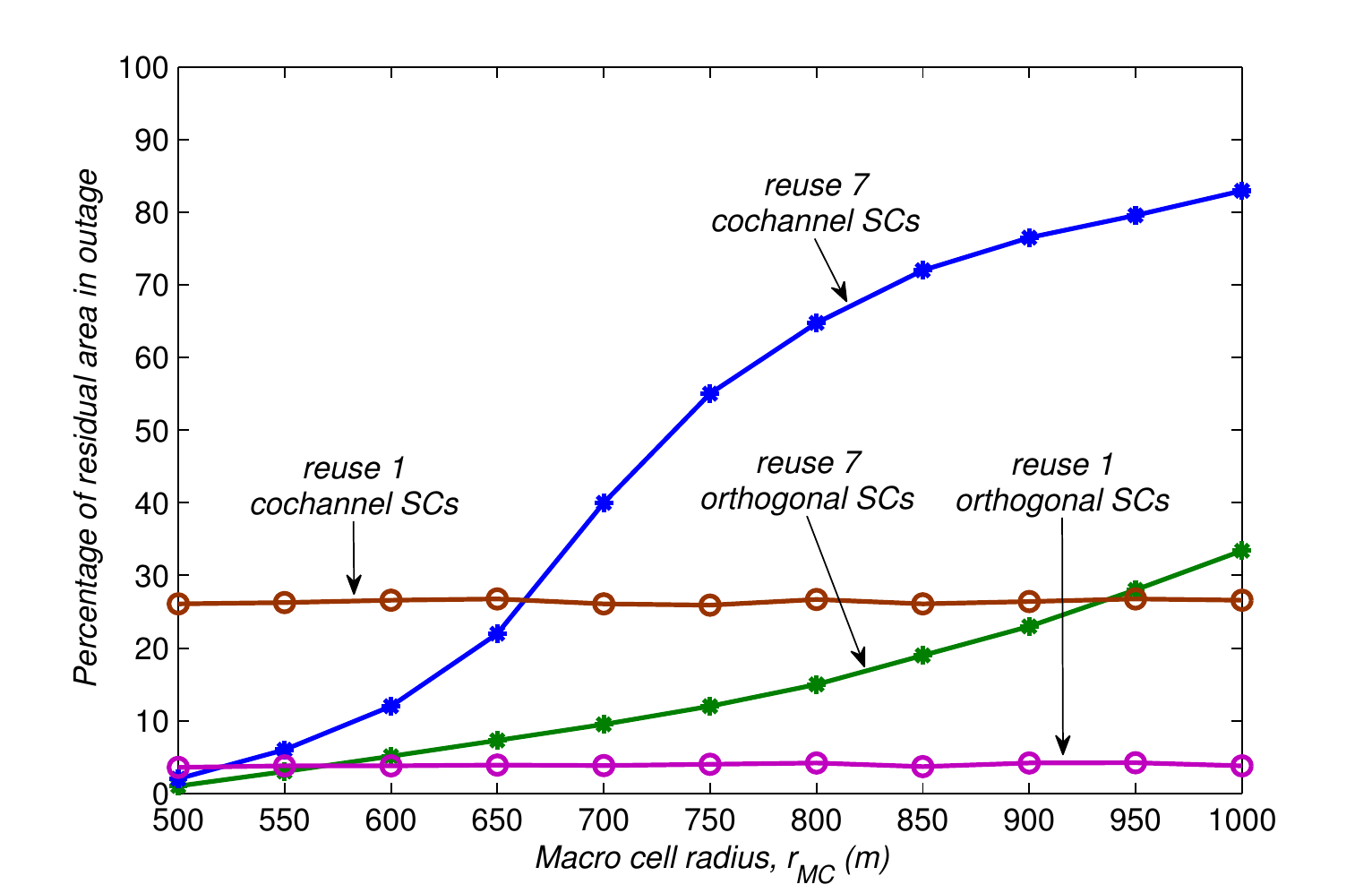}
\caption{The percentage of residual outage area for different values of MC radius in reuse-1 and 7 networks for the target spectral efficiency of $C_{0}= 1 ~\rm{b/s/Hz}$ with $\alpha = 4$}
\label{new_Fig_14}
\end{center}
\end{figure}

The reuse-1 network ends up with a lower residual outage area for moderate to large values of MC radius $r_{\textmd{MC}} \gtrsim 650 ~\rm{m}$ as compared to a reuse-7 network with co-channel SCs ($r_{\textmd{MC}} \gtrsim 600 ~\rm{m}$ with orthogonal SCs). This can be seen in Fig.~\ref{new_Fig_14} where the results for percentage of residual outage area for both reuse-1 and reuse-7 networks for the rate threshold of $C_{0}= 1 ~\rm{b/s/Hz}$, are obtained from simulations. Interestingly, in the range $650 ~\rm{m} \lesssim \emph{r}_{\textmd{MC}} \lesssim 1050 ~\rm{m}$, although the percentage outage area with reuse-7 without SCs is smaller than the one in a reuse-1 network, the reuse-7 network ends up with larger residual outage area in this range even with the use of orthogonal SCs. This is because, in a reuse-7 network, a specific location has to receive at least a SNR of $23.03 ~\rm{dB}$ in order to be in coverage for the target spectral efficiency of $C = (1/7)\log_{2}(1+\texttt{SNR}/\Gamma)=1~\rm{b/s/Hz}$ (this is in compared to $2 ~\rm{dB}$ SIR for the reuse-1 network). Therefore, a large portion of SCs (near edges) are in outage, resulting in a large residual outage area in the reuse-7 network.

\section{Summary and Conclusions}
\label{sec:conc}

In this paper, we analyze the case where small cells are used for range extension and to fill coverage holes in a macrocell network. This leads to a \emph{dependent} placement of SCs within MCs. Specifically, we obtain the average number of required isolated SCs to eliminate coverage holes in the downlink of a reuse-1 macrocell hexagonal-lattice network. When compared with real-world deployments, the presented results provides lower bound on the required number of SCs in cellular networks. This is in contrast to the PPP model which has been shown via simulations to provide an upper bound on the required number of SCs in cellular networks. As a first attempt to analyze dependent placements, the analysis is based on some simplifying assumptions. The key assumptions here are that the MCs are hexagons, that a SC covers a fixed area of $A_{\textmd{SC}}$, and a block model for the shadowing is assumed.

To provide a closed-form analysis, we developed accurate closed-form bounds on the average total interference within a MC. These bounds are particularly simple in the sense that they are expressed as functions of normalized distance to the centre of the MC. The normalization makes the analysis independent of the MC radius and leads to simple analytical expressions for SIR and ROP within MCs. We use these bounds to obtain, for a chosen target spectral efficiency, the fractional area within a MC in outage (with no SCs in use) and the required average number of SCs (per MC) that can completely overcome outage with \emph{isolated} SCs. The tools developed here allow for a quick assessment of the tradeoff between target spectral efficiency, ROP threshold, and required number of SCs.

Furthermore, we show that in an reuse-1 interference-limited network, the percentage of outage area does not depend on either the MC radius or the transmit power. However, since isolated SCs are impractical, when using co-channel or orthogonal SCs, the network experiences a residual outage area which depends heavily on the frequency assignment for SCs. We use simulations to show that, despite placing SCs in coverage holes, with co-channel SCs, the outage area merely moves, but is not reduced. On the other hand, with orthogonal SCs, the outage area in the network is reduced substantially. In particular for the target spectral efficiency of $1~\rm{b/s/Hz}$, the use of orthogonal SCs reduces the outage area by approximately 86\% in a reuse-1 network (further improvements in performance may be achieved by using a resource allocation scheme amongst SCs).

In a fair comparison between the two networks with frequency reuse-1 and reuse-7 for a given target spectral efficiency, the reuse-7 network has less of an outage area for small to moderate values of MC radius $r_{\textmd{MC}}$, but the reuse-1 network achieves better results for moderate to large values of $r_{\textmd{MC}}$. For example, with $r_{\textmd{MC}} =1000 ~\rm{m}$ and the target spectral efficiency of $1~\rm{b/s/Hz}$, the percentage residual outage area in a reuse-1 network with orthogonal SCs is $1/6$ \,of the one in reuse-7 network. The results suggest the use of frequency reuse factor 1 for moderate to large sizes of macrocells.

\bibliographystyle{ieeetr}
\bibliography{ref}
\end{document}